\documentclass[10pt,iop]{emulateapj}
\usepackage{apjfonts}
\usepackage{epsfig}
\usepackage{amsmath}
\begin{document}

\title{The Bright Reverse Shock Emission in the Optical Afterglows of Gamma-ray Bursts in a Stratified Medium}
\author{Shuang-Xi Yi$^{1}$, Xue-Feng Wu$^{2}$, Yuan-Chuan Zou$^{3}$ and Zi-Gao Dai$^{4}$}
\affil{$^{1}$School of Physics and Physical Engineering, Qufu Normal University, Qufu 273165, China; yisx2015@qfnu.edu.cn\\
       $^{2}$Purple Mountain Observatory, Chinese Academy of Sciences, Nanjing 210008, China; \\
       $^{3}$School of Physics, Huazhong University of Science and Technology, Wuhan 430074, China; \\
       $^{4}$School of Astronomy and Space Science, Nanjing University, Nanjing 210093, China.\\}

\begin{abstract}
The reverse shock (RS) model is generally introduced to interpret the optical afterglows
with the rapid rising and decaying, such as the early optical afterglow of GRB 990123 (which is also called optical flash). In
this paper, we collected 11 gamma-ray bursts (GRBs) early optical afterglows, which have such signatures of dominant RS emission. Since the temporal slopes of the optical flashes are determined
by both the medium density distribution index $k$ and the electron spectral index $p$, we apply the RS model of the
thin shell case to the optical flashes and determine the ambient medium of the progenitors. We find that
the $k$ value is in the range of 0 - 1.5. The $k$ value in this paper
is consistent with the result in Yi et al. (2013), where the forward
shock (FS) model was applied to some onset bumps. However, the method adopted in this paper is only applicable to GRB afterglows
with significant sharp rising and decaying RS emission. Our results indicate
that the RS model can also be applied to confirm the circumburst medium, further implying that GRBs may
have diverse circumburst media.
\end{abstract}

\keywords{gamma ray: bursts --- radiation mechanism: non-thermal}

\section{Introduction}
A growing number of gamma-ray burst (GRB) afterglows have been detected with the new observational
detectors since 1997, e.g., the UV-Optical Telescope (UVOT) and X-ray Telescope (XRT) onboard
Swift (Gehrels et al. 2004; Burrows et al. 2005; Roming et al. 2015). According to the fireball model,
GRB afterglows are produced by the
external shock with the surrounding medium through the synchrotron emission (M{\'e}sz{\'a}ros \& Rees 1997;
Wijers et al. 1997; Sari et al. 1998; Piran 1999; Zhang 2007). Different GRB afterglows have greatly improved our understanding on the physical origin of GRBs. Therefore, afterglows provide an important clue
to the nature of the GRB source and ambient medium, especially for the early afterglow, as
it connects the tail of prompt emission with the subsequent afterglow. For example, the observed plateaus and flares in afterglows
imply the long-lasting or restart of the central engine of GRBs (Dai \& Lu 1998a, b; Burrows
et al. 2005; Zhang et al. 2006; Nousek et al. 2006; Dai et al. 2006; Wang \& Dai 2013; Wu et al. 2013;
Yi et al. 2015, 2016, 2017; Si et al. 2018).

Since long GRBs are produced via the core collapse of massive stars, the circumburst medium of long GRBs
may be an interstellar wind. However, Schulze et al. (2011) found that the most of afterglow
light curves were compatible with a constant density medium (ISM case). Only some afterglows show evidence
of an interstellar wind profile at late times. Therefore, we cannot determine the ambient medium types of GRBs.
Besides, there is also no effective method to determine the circumburst medium. Interestingly, Yi et al. (2013) successfully applied
the forward shock (FS) model with a power-law density distribution to 19 onset afterglows and confirmed the circumburst
medium for those GRBs. They found that the value of the ambient medium distribution index $k$ is
in the range of 0.4 - 1.4, with a typical value of $k\sim1$ (also see Liang et al. 2013).
Therefore, GRBs may have different ambient media. In this paper, we apply the reverse shock (RS) model
to the optical flash and determine the circumburst medium types
of GRBs with optical flashes. We use the temporal and spectral indices to determine the values of $k$
and $p$, the method same as Yi et al. (2013), where the FS model was applied to the afterglow onset sample.

The paper is organized as follows. In the section 2, the FS and RS model are applied.
Our optical flash sample selection criteria are presented in Section 3. Next, we investigate 11 optical flash afterglows
in detail. Our conclusions and discussion are presented in Section 5.

\section{Reverse Shock and Forward Shock}

We consider a relativistic fireball shell with an initial kinetic energy $E$
and the initial Lorentz factor $\eta$, which expands into the ambient medium with
the proton number density ${n_1}$. We consider a general power-law density distribution ${n_1} = A{R^{-k}}=n_0(R/R_0)^{-k}$,
where $R$ is the radius, $n_0$=1 $cm^{-3}$ and $R_0$ is the radius at $n_0$. Much work has been done for different types of circumburst medium,
such as, an ISM $k=0$ (Sari et al. 1998; Gao et al. 2013; De Colle et al. 2012), and an interstellar wind $k=2$
(Dai \& Lu 1998c; Chevalier \& Li 2000; Wu et al. 2003, 2005; Zou et al. 2005; De Colle et al. 2012).
During the shell spreading into the environment medium, two shocks are produced:
a RS that propagates into the shell, and a FS that propagates into the ambient
medium (Sari \& Piran 1995; M{\'e}sz{\'a}ros \& Rees 1997; Sari et al. 1998; Kobayashi
2000; Yi et al. 2013, 2014).

The RS emission is mostly discussed for two situations: the thick shell and the thin shell. However, it has been suggested that the observed afterglows could be described well in the thin shell case rather than the thick shell case (Japelj et al. 2014; Gao et al. 2015). So we apply the RS of the thin shell case to the optical light curves in this paper. According to the observations, optical afterglows with the prominent RS emission can be divided into two different types. Type I afterglows show that RS and FS emissions are obvious , and the peak time of RS is earlier than FS emission. Type II afterglows show that the RS component meets the FS emission after the FS peak time, suggesting slow evolution at the late time (Zhang et al. 2003; Gomboc et al. 2009; Harrison \& Kobayashi 2013; Japelj et al. 2014; Gao et al. 2015).

The synchrotron emission from the shocked ambient medium has been discussed in the literature (for more details about the FS emission see  Sari et al. 1998, 1999; Granot \& Sari 2002; Yi et al. 2013; Gao et al. 2013; Zhou et al. 2020).
For the shocked fireball shell emission in the thin shell case, two characteristic frequencies and peak flux density of the RS at the crossing time $T_{\Delta}$ for $k=0$ are (Kobayashi 2000; Yi et al. 2013)
\begin{equation}
{\nu^{r}_{m,\Delta}} = 2.16\times 10^{12} \varepsilon _{e,r,-1}^{2}\,\varepsilon _{B,r,-1}^{1/2} n_{0}^{1/2} \eta_2^{2}\,(1+z)^{-1}\,{\rm Hz},
\end{equation}
\begin{equation}
{\nu^{r}_{c,\Delta}} = 1.80\times 10^{15} \varepsilon _{B,r,-1}^{-3/2}(1+Y_r)^{-2}E_{52}^{-2/3} n_{0}^{-5/6} \eta_2^{4/3}\,(1+z)^{-1}\,{\rm Hz},
\end{equation}
and
\begin{equation}
F_{\nu ,\max, \Delta }^{RS} =  7.39 \varepsilon _{B,r,-1}^{1/2}E_{52} n_{0}^{1/2} D_{L,28}^{-2} \eta_2 \,(1+z)\,{\rm Jy},
\end{equation}
where the Compton parameter $Y_r=1$, $E=10^{52} \,erg$, $\eta=100$, $n_0=1$ $cm^{-3}$, $D_L=10^{28} \,cm$, $p=2.5$, and $\epsilon _{B,r}=\epsilon _{e,r}=0.1$ are applied.
The scalings laws before and after the reverse shock crossing time are\footnote{Considering the high pressure for the shocked shell after the RS crossing time, the evolution of the shocked shell is dominated by adiabatic expansion. Therefore, the evolution of $\nu_c$ ($\gamma_c$) is considered to be the same as $\nu_m$ ($\gamma_m$) for both thick shell and thin shell cases after the RS crossing the shell, which is different from Kobayashi (2000). }
\begin{equation}
T < {T_\Delta }:\,\,{\nu^{r}_m} \propto {T^{6}},{\nu^{r}_c}
\propto {T^{-2}},F_{\nu ,\max }^{RS} \propto
{T^{\frac{{3}}{{2}}}},
\end{equation}
and
\begin{equation}
T > {T_\Delta }:\,\,{\nu^{r}_m} \propto {T^{ - \frac{73}{48}}},{\nu^{r}_c}
\propto {T^{- \frac{73}{48}}},F_{\nu ,\max }^{RS}
\propto {T^{- \frac{47}{48}}}.
\end{equation}
The different theoretical light curves of RS emission are shown in Figs. 1 and 2.

The FS model predicts a smooth early onset bump in the afterglow light
curves when the fireball shell is decelerated by the environment medium
(Sari et al. 1998; Molinari et al. 2007; Liang et al. 2010).
Such an onset feature generally has a smooth rising and decaying slopes.
The rising index is generally less than 3, and mostly between 1 and 2 in
a statistical sense (Liang et al. 2010). Therefore, only some observed onset bumps are consistent with
the theoretical expectations in the ISM case. From Eqs. 67 and 68
of Yi et al. (2013), the values $k$ and $p$ are determined by the rising and decaying slopes
of the FS. This implies that different slopes of light
curves may have different circumburst medium ($k$).

From the above discussion, we know that the FS model cannot explain the light curves with the rising
index larger than 3. Therefore, in this paper, we
apply the RS model to the optical light curves with rapidly rising and decaying, and determine the type of circumburst
medium of GRBs with optical flashes. In a general power-law density distribution ${n_1} = A{R^{-k}}$, the RS theoretical flux density before and after the crossing time are:
\begin{equation}
  \label{eq:A_k_nu}
  F_\nu ^{RS} (T<T_{\Delta})\propto \left\{
  \begin{array}{ll}
   {T^{ \frac{{12p - 5kp + 4k - 10}}{4}}}{\nu ^{ - \frac{p}{2}}},\quad \nu
        > \max\left\{ {\nu _c^r,\nu _m^r} \right\} \\
   {T^{\frac{{12p - 5kp + k - 6}}{4}}}{\nu ^{ - \frac{{p - 1}}{2}}},\quad
        \nu _m^r < \nu  < \nu _c^r\\
   {T^{\frac{{2 - k}}{4}}}\;{\nu ^{ - \frac{1}{2}}}\;,\quad \quad
       \;\nu _c^r < \nu  < \nu _m^r
  \end{array}
  \right.
\end{equation}
and
\begin{equation}
  F_\nu ^{RS}(T>T_{\Delta})\propto \left\{
  \begin{array}{ll}
   {T^{ - \frac{{4 + p}}{2}}}{\nu ^{ - \frac{p}{2}}},\quad \nu  >
       \max\left\{{\nu _c^r,\nu _m^r} \right\}\\
   {T^{ \frac{{14kp + 6k - 73p - 21}}{{24(4 - k)}}}}{\nu ^{ - \frac{{p
      - 1}}{2}}},\quad \nu _m^r < \nu  < \nu _c^r\\
   {T^{ \frac{{34k - 167}}{{24(4 - k)}}}}\;{\nu ^{ - \frac{1}{2}}}\;.\quad \quad \;\nu
      _c^r < \nu  < \nu _m^r
  \end{array}
  \right.
\end{equation}
As shown in Eqs. 6 and 7, the temporal slopes before and after the crossing time are determined
by both the medium density profile $k$ and the electron spectral index $p$. Therefore, the types of the ambient medium
density profile and the distribution of the electrons affect the RS rise and decay indexes.

To understand the nature of RS and FS emissions, we compare the typical parameters in both shocks.
Since theoretical models predict that mildly magnetized
baryonic outflows of GRBs will produce strong RS emission (Zhang \& Kobayashi 2005;
Gomboc et al. 2009). In order to produce an obvious optical flash, there should be a strong
RS emission that outshines the FS component, especially at the early time. Thus we can derive
the ratio of characteristic frequencies and peak flux density of RS ($p, \varepsilon _{B, r},
\varepsilon _{e, r}$) and FS ($p, \varepsilon _{B, f}, \varepsilon _{e, f}$), which is also
seen in Kobayashi \& Zhang (2003) and Zhang et al. (2003),
\begin{equation}
{\nu^{r}_{m,\Delta}/\nu^{f}_{m,\Delta}} = \left[\frac{9^{12-4k} (3-k)^{24-8k}}{2^{12-4k}7^{24-8k}}\right]^{1/2(3-k)}\,R_B\,R_{e}^2\,\eta^{-2},
\end{equation}
\begin{equation}
{\nu^{r}_{c,\Delta}/\nu^{f}_{c,\Delta}} = R_{B}^{-3},
\end{equation}
and
\begin{equation}
{F_{\nu ,\max, \Delta }^{RS}/F_{\nu ,\max, \Delta }^{FS}} = \left[\frac{2^{18-6k} 7^{6-2k}}{9^{6-2k}(3-k)^{6-2k}}\right]^{1/2(3-k)} \,R_B\,\eta,
\end{equation}
where $R_B=(\varepsilon _{B, r}/\varepsilon _{B, f})^{1/2}$, $R_e=\varepsilon _{e, r}/\varepsilon _{e, f}$, and
$\eta$ is the initial Lorentz factor.

\section{Optical Sample Selection and Light Curve Fitting}

Due to small durations of short GRBs, their very early afterglow emissions are hardly detected.
Consequently, the reverse shock emission with rapid rising and decaying features rarely appears at the optical
band for short GRBs. However, observations of radio afterglows show a reverse shock feature from some short and long GRBs
(Lloyd-Ronning et al. 2017; Lloyd-Ronning 2018; Lamb et al. 2019), although the rise was not observed. Since short GRBs are usually found to arise from binary compact star mergers, their circumburst medium is generally considered as a constant density medium. Hence, afterglows from short GRBs are excluded in our sample. Furthermore, because of peaks at very late times, we don't consider radio afterglows to constrain the circumburst medium of GRBs.

Many early optical afterglow light curves have been detected. It has been expected that
different optical afterglows may have different emission components (Rykoff et al. 2009; Kann et al. 2010; Liang et al. 2010; Panaitescu, \& Vestrand 2011; Li et al. 2012). As the FS model
cannot explain the light curves with the rapidly rising and decaying, the RS model is
usually applied to interpret those afterglows, such as the optical afterglow of GRB 990123,
which is due to the emission of a RS propagating into the fireball shell (Sari \& Prian 1999a; Kobayashi 2000, 2003; Wu et al. 2003;
Zhang et al. 2003; Harrison \& Kobayashi 2013). Recently, some GRBs with the RS signatures are collected and discussed for the RS emission in a constant ISM environment (Japelj et al. 2014; Gao et al. 2015). In fact, the rising and decaying slopes are used to determine
the values of $k$ and $p$ (Eqs. 6 and 7 for RS in this paper; Eqs. 67 and 68 for FS of Yi et al. 2013 ),
therefore, different environment media may have different slopes of light curves.
Considering only the ISM case, it may not be the real conditions for different GRBs.

In this paper, we widely search for the peculiar features of the GRB afterglows and two criteria are employed:
(1) a remarkable rising and decaying feature appears in an optical afterglow light curve;
(2) the rising slope is generally larger than 3.
We select 11 optical afterglows at an early time, which have such features in dominant RS emission. We
find that there is a remarkable FS signature after the peak of the light
curve. Thus, we use the RS plus FS model to fit the 11 light curves. The
fitting results are reported in Fig. 3 and Table 1.

11 optical light curves dominated by RS emission at early times, roughly
speaking, could be divided into two different types. GRB 041219A is regarded as
type I and the other sources belong to type II.
Theoretical models predict that mildly magnetized baryonic outflows of GRBs produce
strong reverse shock emission (Zhang \& Kobayashi 2005; Gomboc et al. 2009). Therefore,
to produce type I and type II optical light curves, there should be a strong RS
emission that outshines the FS component, especially at early times, the ratio
parameters $R_{B}>1$ and $F_{\nu ,\max, \Delta }^{RS}/F_{\nu ,\max, \Delta }^{FS}>1$
are required. Table 2 shows the ratio of RS and FS parameters. Our results show
that if a remarkable RS component appears in an optical afterglow, $\varepsilon _{B,r}$
should be larger than $\varepsilon _{B,f}$. The larger $\varepsilon _{B,r}$/
$\varepsilon _{B,f}$, the more obvious RS component. Our results are also
consistent with those of Japelj et al. (2014) and Gao et al. (2015).

\section{Case Study}
We apply the RS model to the optical flashes and confirm
the circumburst medium of the selected GRBs. However, since a remarkable FS signature gradually appears
after the peak of the light curve, we also use the FS model to fit the late evolution parts of
light curves. Because of the pollution of the FS emission after the peak time of the flash, we adopt the
rising slope, rather than the decaying slope, connecting with the optical spectrum to determine
the environment medium of GRBs for different emission regimes (see Eqs. 6 and 7).
We suppose it is reliable to determine the ambient medium distribution index ($k$) using the temporal
slopes and spectral behavior. Here we suppose that a
power law form for the energy distribution of the shocked electrons with the Lorentz factor of electrons
larger than the minimum Lorentz factor, as introduced in Sari et al. (1998), to keep the energy of the
electrons to be finite, so the index $2<p<3$ is supposed in this paper. Please also see the description of a power-law accelerated electron in Higgins. et al. (2019). In our numerical fit, the least-$\chi^2$ method is applied to those optical afterglows. However,
since those optical light curves in our paper are composed of two or more power-law segments along with some
erratic pulses, jet breaks, plateaus or rebrightening features and so on, it is hard to fit well for the
whole light curve of each GRB. Here, we make our fits only around the rising and decaying phases when using
both RS + FS emission, and exclude the mixed components such as erratic pulses, jet breaks, plateaus, or
rebrightening features. Because of the poor limit conditions, and many parameters
of the fitting models (RS and FS), we cannot confirm all the parameters exactly except for $k$ and
$p$. Therefore we attempt to give reasonable fitting parameters for each GRB. The results
are shown in Table. 1. The case of $\nu < \left\{ {\nu _{c, r},\nu _{m, r}} \right\}$
is not considered, because it is almost unlikely in optical spectra. We ignore synchrotron
self-absorption in the optical emission, since it is unimportant at most times for
optical afterglows (also see Japelj et al. 2014).

\subsection{GRB 990123}

GRB 990123 is one of the brightest GRBs detected by {\em BeppoSAX}. This famous GRB was intensively
studied by many authors, who generally supposed that this optical flash should be attributed to the RS
emission (Castro-Tirado et al. 1999; Sari \& Piran 1999b; Dai \& Lu 1999; Wang et al. 2000; Fan
et al. 2002; Panaitescu \& Kumar 2004; Nakar et al. 2005). Some authors considered the circumburst
medium being an ISM case (Fan et al. 2002; Nakar et al. 2005; Japeij et al. 2014), and applied the RS
model in the ISM case to fit the optical flash. In our study, we fit the optical flash with an empirical smooth broken
power-law function, and the rising index of GRB 990123 is about $\alpha_1=3.31\pm0.25$.
However, the theoretical rising slope of the RS is 5 or 6 in the ISM case, much steeper than the real value
of GRB 990123. We take the value of $k$ to be free and apply the
RS model of this paper to this case, and then there are two possible values for the power-law index
of the energy distribution with the optical spectral index $\beta_o=0.60\pm0.04$ (Maiorano et al. 2005).
In the first case, the value $p=2\beta_o=1.20\pm0.08$ for $\nu > \max\left\{ {\nu _c^r,\nu _m^r} \right\}$. The
value of $p$ is not in a reasonable range (we suppose $2 < p < 3$ ), therefore the model of
$\nu > \max\left\{ {\nu _c^r,\nu _m^r} \right\}$ is unsuitable for GRB 990123. In the second case,
the value $p=2\beta_o+1=2.20\pm0.08$ for $\nu _m^r < \nu  < \nu _c^r$, and the rising index is $\alpha_1=(12p-5kp+k-6)/4$,
therefore $k=(4\alpha_1+6-12p)/(1-5p)=0.71\pm0.12$. The values of $k$ and $p$ are both in their reasonable
ranges, so we suggest that the circumburst medium of GRB 990123 could not be an ISM case and determine the value of
$k=0.71$.

\subsection{GRB 041219A}

GRB 041219A is an extraordinary GRB and thought to produce emission from two different physical
processes, the internal emission from the central engine of GRB itself and external RS - FS emission that are
emerged in the very early optical and near-infrared emission (Blake et al. 2005; Fan et al. 2005;
G{\"o}tz et al. 2011). This is also the only one afterglow regarded as type I in this paper, both
showing obvious RS and FS light curve peaks. Because of the lack of the optical spectrum for this GRB, according
to the analysis of Blake et al. (2005) and Fan et al. (2005), we apply $p=2.4$ with the emission
regime $\nu _m^r < \nu  < \nu _c^r$, and confirm $k=(4\alpha_1+6-12p)/(1-5p)=0.80\pm0.10$
with the rising slope $\alpha_1=3.50\pm0.21$.

\subsection{GRBs 060607A and 061007}

GRBs 060607A and 061007 are long and fairly hard GRBs detected by {\em Swift}/BAT. Some authors considered the
two GRBs as the afterglow onsets and determined the initial Lorentz factor with the peak time and isotropic
energy (Molinari et al. 2007; Jin \& Fan 2007; Ziaeepour et al. 2008; Covino et al. 2008; Liang et al. 2010).
However, the optical light curves of GRBs 060607A and 061007 have very sharp rising $T^{3.21\pm 0.11}$ and
$T^{3.93\pm 0.03}$, respectively, which is inconsistent with the FS model. For the $\nu > \max\left\{ {\nu _c^r,\nu _m^r} \right\}$
case, is not suitable for GRB 060607A, since $p=2\beta_o=1.12\pm0.08$ with the optical spectral index
$\beta_o=0.56\pm0.04$ (Nysewander et al. 2009). While $p=2\beta_o+1=2.12\pm0.08$ for $\nu _m^r < \nu  < \nu _c^r$,
and we also determine the value of $k=(4\alpha_1+6-12p)/(1-5p)=0.69\pm0.08$. For the GRB 061007,
the value $p=2\beta_o=1.56\pm0.04$ with the optical spectral index $\beta_o=0.78\pm0.02$ (Zafar et al. 2011) for
$\nu > \max\left\{ {\nu _c^r,\nu _m^r} \right\}$, while $p=2\beta_o+1=2.56\pm0.04$ and
$k=(4\alpha_1+6-12p)/(1-5p)=0.76\pm0.03$ for $\nu _m^r < \nu  < \nu _c^r$.

\subsection{GRB 081007}

According to Jin et al. (2013), the optical flash of GRB 081007 has rising and decaying indices of about 3
and $-2$ respectively, followed by the broken power-law lines with the slope indices of $-0.65$ and $-1.25$. The X-ray afterglow
is nearly the same as the optical light curve after 300 s. By using the RS and FS model to fit the
light curve, we find the RS model can fit the optical flash but the FS emission is inconsistent with
the afterglow after the peak time of the flash. Since this GRB presents a transition to a shallow decay phase with the decay index
of about 0.60, the standard FS model cannot account for the shallow decay phase dominated by the continued energy injection
(Dai \& Lu 1998a, b; Zhang \& M\'esz\'aros 2001; Dai 2004; Zhang 2013; L{\"u} \& Zhang 2014; Hou et al. 2014),
which is also seen the fit result of GRB 081007 from Japelj et al. (2014), they also used the single RS emission to the optical afterglow of the early time. Therefore, we only use the RS model
to fit the optical flash, and the rising index of GRB 081007 is $3.01\pm0.07$.
We exclude the emission regime of $\nu > \max\left\{ {\nu _c^r,\nu _m^r} \right\}$ for the optical index
$\beta_o=0.86\pm0.07$ (Japelj et al. 2014). We find $p=2\beta_o+1=2.72\pm0.14$ and
$k=(4\alpha_1+6-12p)/(1-5p)=1.16\pm0.07$ for $\nu _m^r < \nu  < \nu _c^r$.

\subsection{GRB 081008}

GRB 081008 is also one of the GRBs dominated by the RS at early times in this paper, in which case the rising index
$\alpha_1=3.38\pm0.12$ (Yuan et al. 2010). We use the two different conditions of Eqs. (6) and (7) to confirm
the values $k$ and $p$ with the optical spectral index $\beta_o=1.10\pm0.04$ (Yuan et al. 2010). For $\nu _m^r < \nu  < \nu _c^r$,
the index $p=2\beta_o+1=3.20\pm0.08$. As mentioned above, we mainly consider the range for $2< p <3$, and therefore, we do not
apply this emission regime to explain GRB 081008. Then the indices $p=2\beta_o=2.20\pm0.08$ and $k=(4\alpha_1+10-12p)/(4-5p)=0.41\pm0.15$
are obtained for $\nu > \max\left\{ {\nu _c^r,\nu _m^r} \right\}$.

\subsection{GRB 090102}

The optical light curve of GRB 090102 is basically similar to the optical afterglow of GRB 990123, which has a sharp rise and
decay evolution after the peak of light curve. The observed optical data are taken
from Klotz et al. (2009) and Gendre et al. (2010). The value of $p$ is not suitable for $\nu > \max\left\{
{\nu _c^r,\nu _m^r} \right\}$ with $\beta_o=0.93\pm0.18$, and therefore, we obtain $p=2\beta_o+1=2.86\pm0.36$
and $k=(4\alpha_1+6-12p)/(1-5p)=1.15\pm0.21$ with $\alpha_1=3.21\pm0.40$ for $\nu _m^r < \nu  < \nu _c^r$.

\subsection{GRB 110205A}

GRB 110205A was a very bright burst detected by {\em Swift}/BAT. The optical data also show the various radiation
features expected in the fireball model, internal emission, external RS and FS emission (Gendre et al. 2012).
There is a very obvious internal component in the early time of the optical light curve (see the Fig. 1 of Gendre et al. 2012),
then we apply the RS - FS model after 400 s. The rising index $\alpha_1=3.88\pm0.08$ and the optical spectral
index $\beta_o=0.84\pm0.04$ (Gendre et al. 2012). Considering a reasonable range of $p$, we apply the
$\nu _m^r < \nu  < \nu _c^r$ to the GRB, and determine $p=2.68\pm0.08$ and $k=0.86\pm0.06$.

\subsection{GRB 130427A}

GRB 130427A was detected by several space instruments, which is a bright and energetic GRB followed by
the multi-wavelength afterglow including the GeV emission (Tam et al. 2013; Liu et al. 2013;
Fan et al. 2013; Kouveliotou et al. 2013; Perley et al. 2014; Vestrand et al. 2014; Ackermann et al. 2014). GRB 130427A was also
found to be accompanied by bright, broad-lined Type Ic supernovae (SN 2013cq) (Xu et al. 2014). There is
an obvious optical flash detected at very early times, and described well by RS and FS emissions with an interstellar
wind environment ($k=2$) for the thin-shell model (Laskar et al. 2013; Perley et al. 2014; Vestrand et al. 2014),
while some authors also applied the ISM case to describe the optical afterglow (Panaitescu et al. 2013;
Maselli et al. 2014). We use our model to confirm the type of environment with the rising index
$\alpha_1=1.63\pm0.10$ and the optical spectral index $\beta_o=0.70\pm0.05$ (Vestrand et al. 2014).
We obtain $p=2\beta_o+1=2.40\pm0.10$ for $\nu _m^r < \nu  < \nu _c^r$, and obtain $k=1.48\pm0.06$, which
is almost the same as Kouveliotou et al. (2013).

\subsection{GRB 140512A}

The optical afterglow of GRB 140512A was detected by 0.8 m TNT in white and R bands
started at $T_0 + 126 $ s after the BAT trigger (Huang et al. 2016). Huang et al. (2016)
reported the observations of a very bright optical lightcurve. The RS-dominated optical emission of
GRB 140512A at early times with the rising index $\alpha_1=3.04\pm0.09$ and the spectral index $\beta_o=0.86\pm0.01$.
Therefore, the electron spectral index $p=2\beta_o+1=2.72\pm0.02$ in the reasonable range,
and $k=(4\alpha_1+6-12p)/(1-5p)=1.15\pm0.03$ for $\nu _m^r < \nu  < \nu _c^r$.

\subsection{GRB 161023A}

GRB 161023A was an exceptionally luminous GRB with redshift z = 2.710, and its afterglow reached a peak
magnitude of r= 12.6 mag (de Ugarte Postigo et al. 2018). The first four data points show the sharp temporal before 100 s,
which are likely the tail of the gamma-ray emission, and behave in a way similar to the prompt emission of GRB 161023A.
The rising slope of optical lightcurve is too steep to be consistent with the expected values for an FS. Therefore,
it is possible that the early optical peak is dominated by the RS emission.
The optical spectral slope $\beta_o=0.66\pm0.08$ and the rising slope $\alpha_1=5.44\pm0.13$ of GRB 161023A, and then the electron spectral index $p=2\beta_o+1=2.32\pm0.16$ and $k=(4\alpha_1+6-12p)/(1-5p)=0.01\pm0.19$ for $\nu _m^r < \nu  < \nu _c^r$ case.
The slope of the density profile $k =0.01\pm0.19$, implying the circumburst environment of GRB 161023A may be a homogeneous interstellar medium, which is consistent well with the theoretical afterglow model in an ISM in the thin shell case (Kobayashi 2000; Yi et al. 2013).

We apply the RS model with
the uncertain density index $k$ to the optical sample, and determine $k$ in the range of $0 - 1.5$.
Fig. 4 shows the distributions of the characteristics of $k$ and $p$. In this figure, the blue dash-dotted lines are the
results determined by the RS, and the data of the red dashed lines are taken from Yi et al. (2013).
We find the distribution of $k$ determined by the RS is consistent with the results confirmed by the FS model,
implying this is an effective method to constrain the environment medium and GRBs may have
different circumburst media.

\section{Conclusions and Discussion}
GRBs are generally believed to be produced from relativistic jets, even though the components
of the jets are magnetized or standard baryonic materials that have been not well confirmed. If a highly magnetized jet appears in a GRB ($\sigma\gg1$) (Giannios et al. 2008), the RS emission would be significantly suppressed or disappeared.
However, the afterglows with onset features are consistent with the standard baryonic fireball
model well (where $\sigma<1$). Therefore, for simplicity, we assume the jet is mildly magnetized. In fact,
our results also show that Type I and Type II optical afterglows would have
mildly magnetized baryonic jets. There should be a strong RS emission that outshines
the FS component. Especially at the early time, the ratio parameters $R_{B}>1$ and
$F_{\nu ,\max, \Delta }^{RS}/F_{\nu ,\max, \Delta }^{FS}>1$ are required.

We have extensively searched for the optical afterglows with the signatures of dominant RS emission
from early afterglow light curves, and 11 GRBs are collected. The RS plus FS model was applied to fit
the selected sample. We found the distribution of $k$ determined by the RS model are in the range of $0 - 1.5$, which is consistent with the results confirmed by the FS emission
from Yi et al. (2013). The RS model was also used to determine
the circumburst medium of GRBs. The ambient density distribution usually follows $R^{-2}$ for the wind environment,
as discussed that long GRBs are associated with core-collapse of the massive stars. However, the results of this paper, connecting with the consequence of Yi et al.
(2013), reveal that the circumburst medium of GRBs may have different forms between an ISM and an
interstellar wind. It implies that the progenitors of some long GRBs may experience a different mass-loss
evolution.

We found that the values of k are concentrated between 0 and 2, peaking at around 1. We notice
this result is shown in both the early optical afterglows (this work) and in the late
afterglows (Yi et al. 2013). This indicates it might be a common feature, the
circumburst environment is neither an ISM nor a wind. The environment is inconsistent
with the late evolution of a massive star.  At the end of a massive star, the nuclear burning
becomes more and more violent, and the stellar wind becomes more massive and faster. It
turns out a much steeper slope of the environment density. On the contrary, the
environment may indicate a binary merger origin, of which one is a neutron star or a black
hole and the other is a white dwarf or a CO-core star (Rueda et al. 2018). To identify this
scenario, gravitational waves might be a unique tool. As the gravitational wave frequency from merging may not
lie on the LIGO's kHz band, the future lower frequency detectors such as
DECIGO (Kawamura et al. 2008), Tianqin (Luo et al. 2016), and LISA (Amaro-Seoane et al. 2017)
might be ideal instruments to identify the single origin or binary origin of
long GRBs. Considering the variety of GRBs, there might be both origins.

\section*{Acknowledgments}
We would like to thank an anonymous referee for valuable comments.
We also thank Ruo-Yu Liu and En-Wei Liang for useful comments and help.
This work is supported by the National Natural Science Foundation
of China (Grant Nos. 11703015, 11725314, U1738132, 11833003), the National Key Research and Development Program of China (Grant No. 2017YFA0402600) and the Natural Science Foundation of Shandong Province (Grant No. ZR2017BA006).

\clearpage
\begin{figure}
  \begin{center}
  \centerline{ \hbox{ \epsfig
  {figure=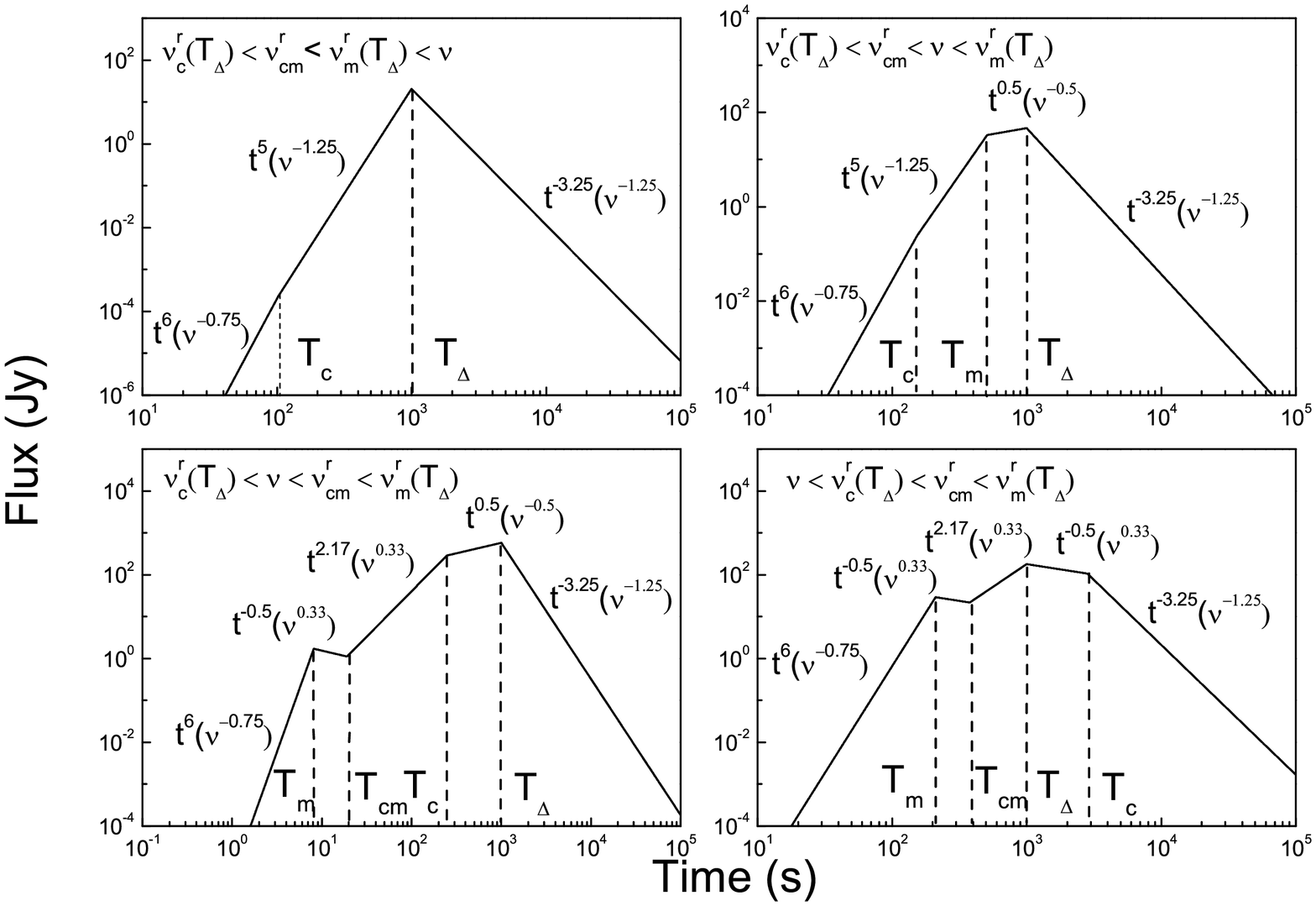,width=5.5in,height=3.8in,angle=0}}}
\caption{Characteristic light curves of a reverse shock in the thin-shell fast-cooling
regime for $k=0$. $T_c$ and $T_m$ are the typical times when $\nu_c$ and
$\nu_m$ cross the observing frequency $\nu$, respectively. $T_{\rm
cm}$ is the time when $\nu_c=\nu_m$. The parameters of light curves
are $E=10^{54}$ erg, $\eta=100$, $z=1$, $p=2.5$,
and $\epsilon _{B,r}=\epsilon _{e,r}=0.4$. In order to show the light curves, we apply
the crossing time $T_{\Delta}=1000$ s.}
  \end{center}
  \end{figure}

\begin{figure}
  \begin{center}
  \centerline{ \hbox{ \epsfig
  {figure=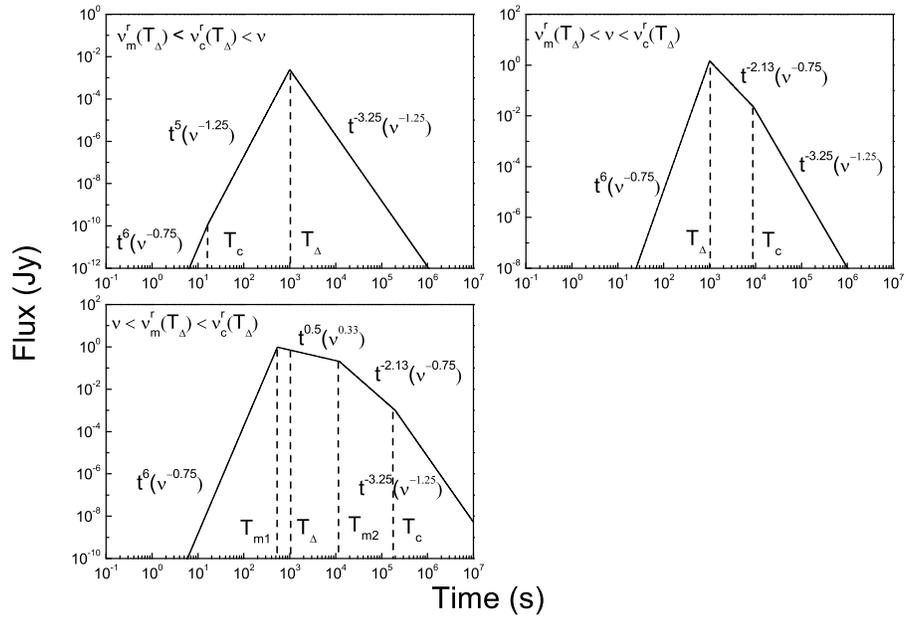,width=5.5in,height=3.8in,angle=0}}}
\caption{Characteristic light curves of a reverse shock in the thin-shell slow-cooling
regime for $k=0$. The parameters of light curves
are $E=10^{52}$ erg, $\eta=100$, $z=1$, $p=2.5$,
and $\epsilon _{B,r}=\epsilon _{e,r}=0.1.$}
  \end{center}
  \end{figure}


\newpage
\begin{figure*}
\includegraphics[angle=0,scale=0.3]{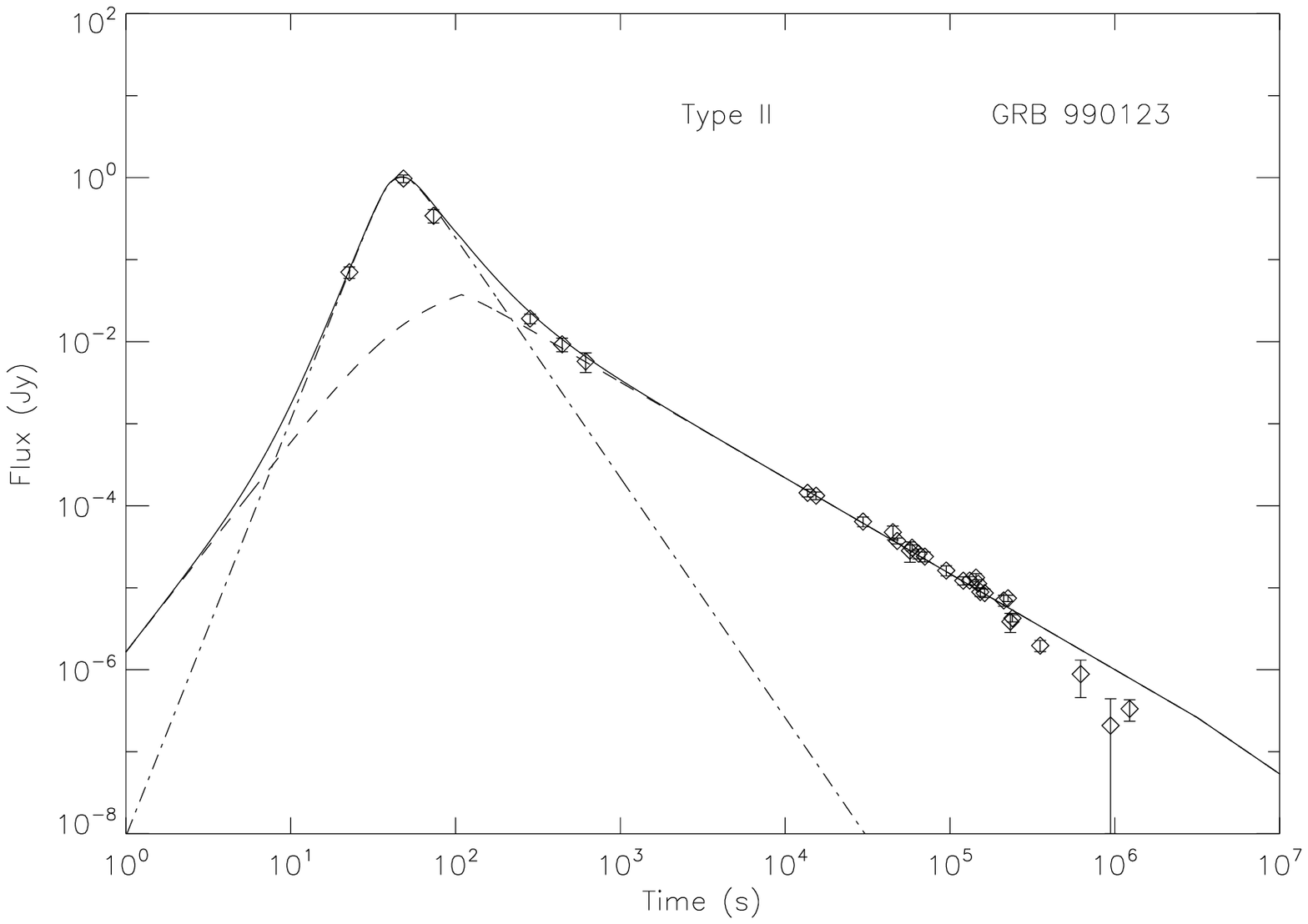}
\includegraphics[angle=0,scale=0.3]{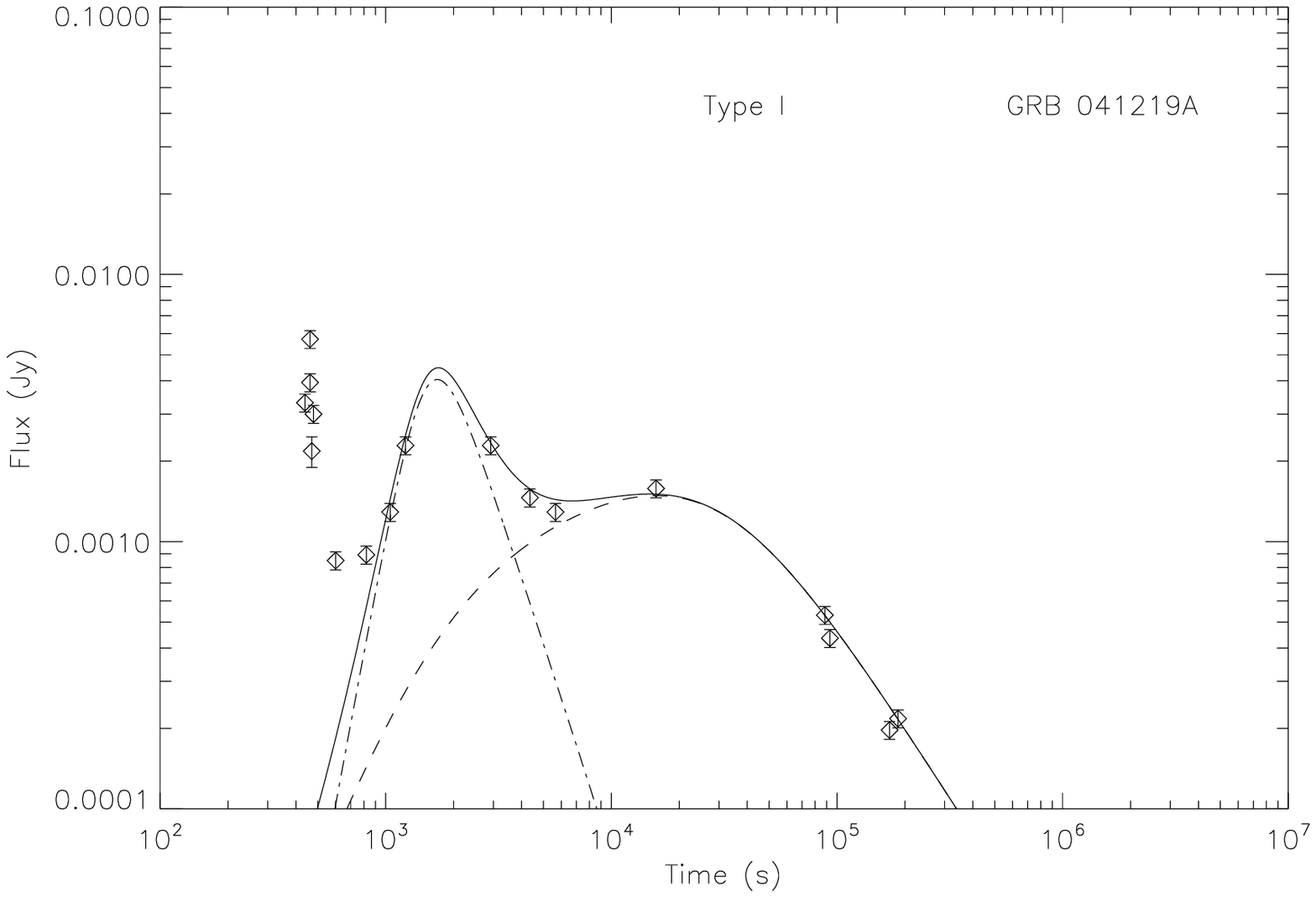}
\includegraphics[angle=0,scale=0.3]{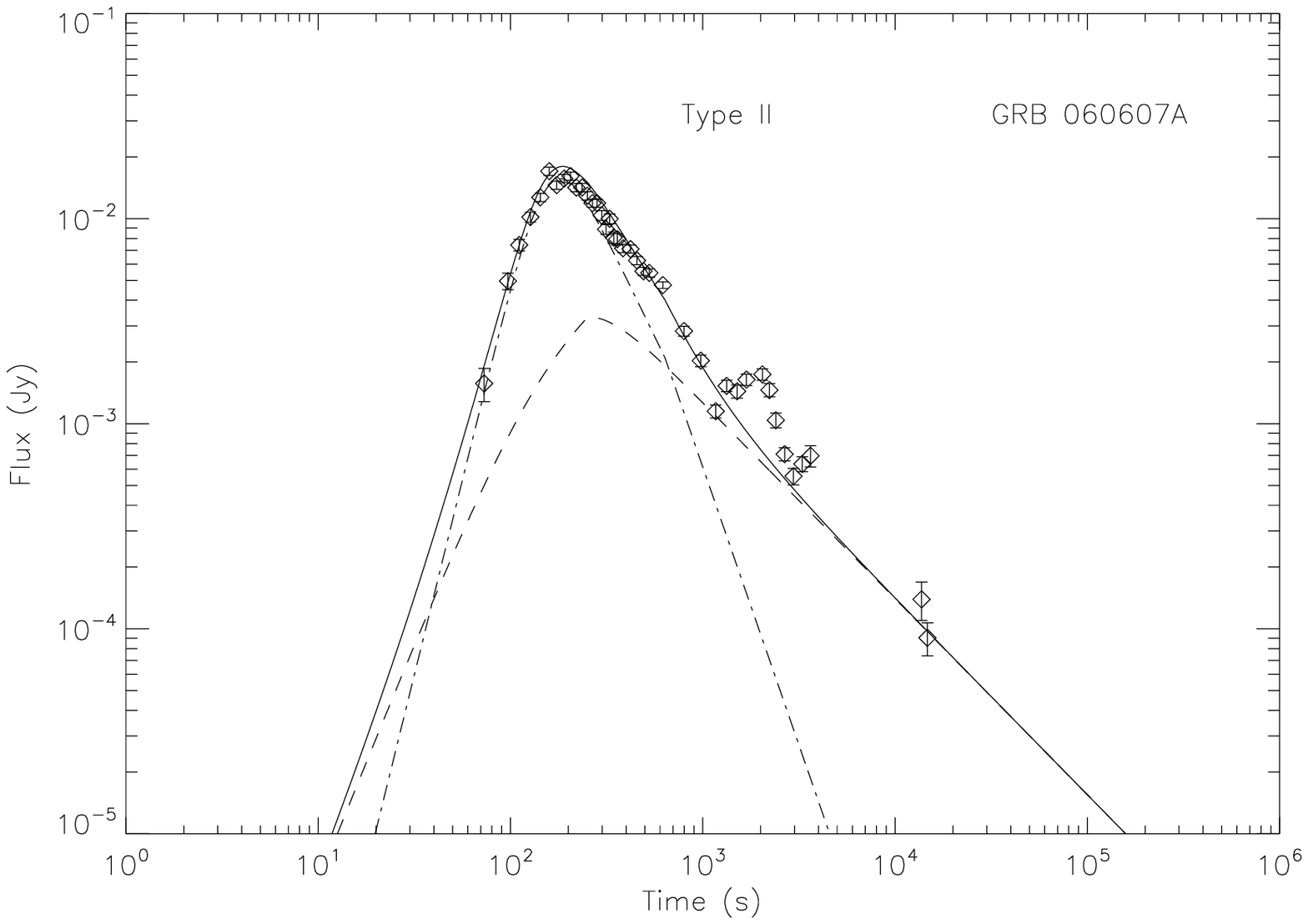}
\includegraphics[angle=0,scale=0.3]{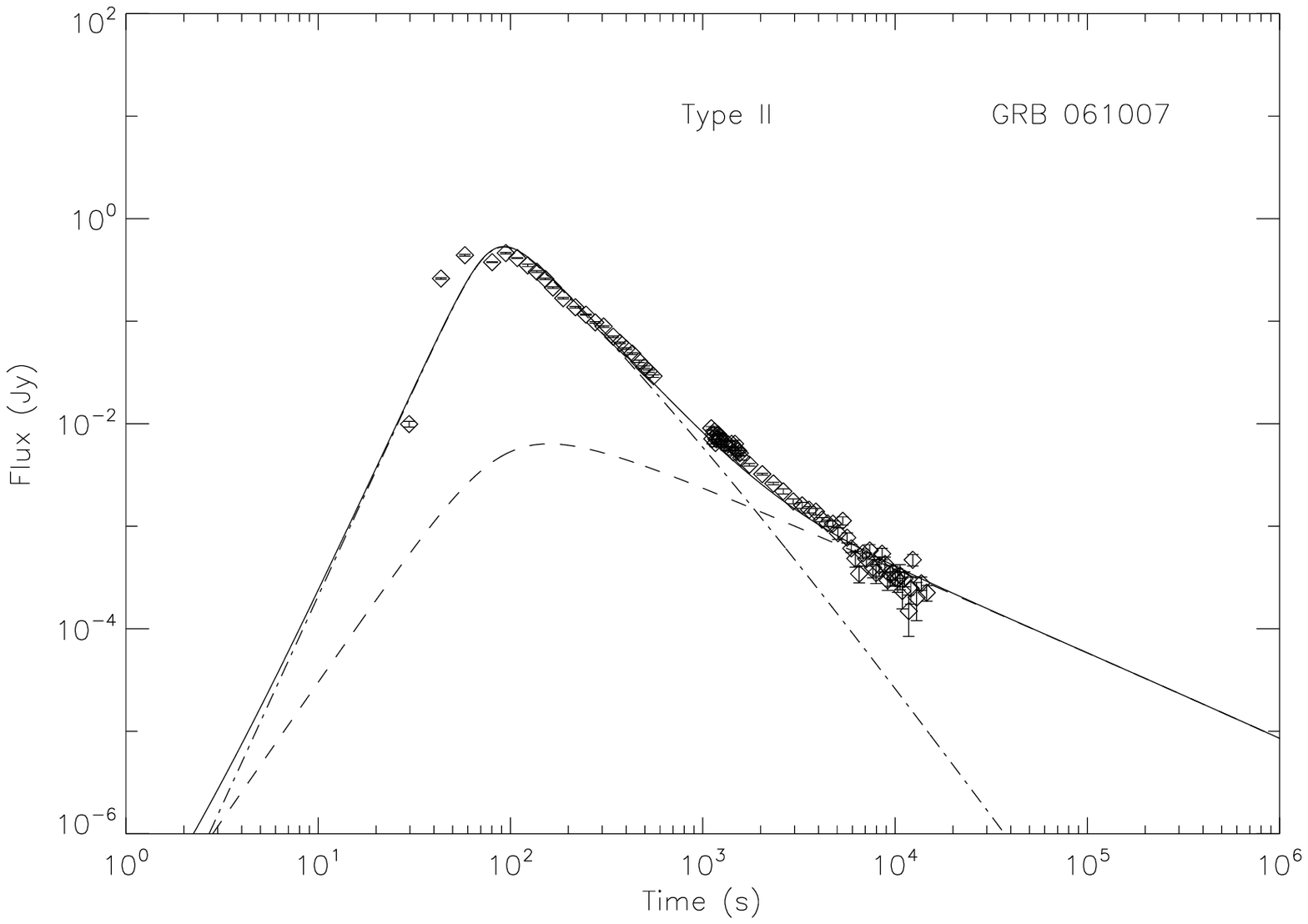}
\includegraphics[angle=0,scale=0.3]{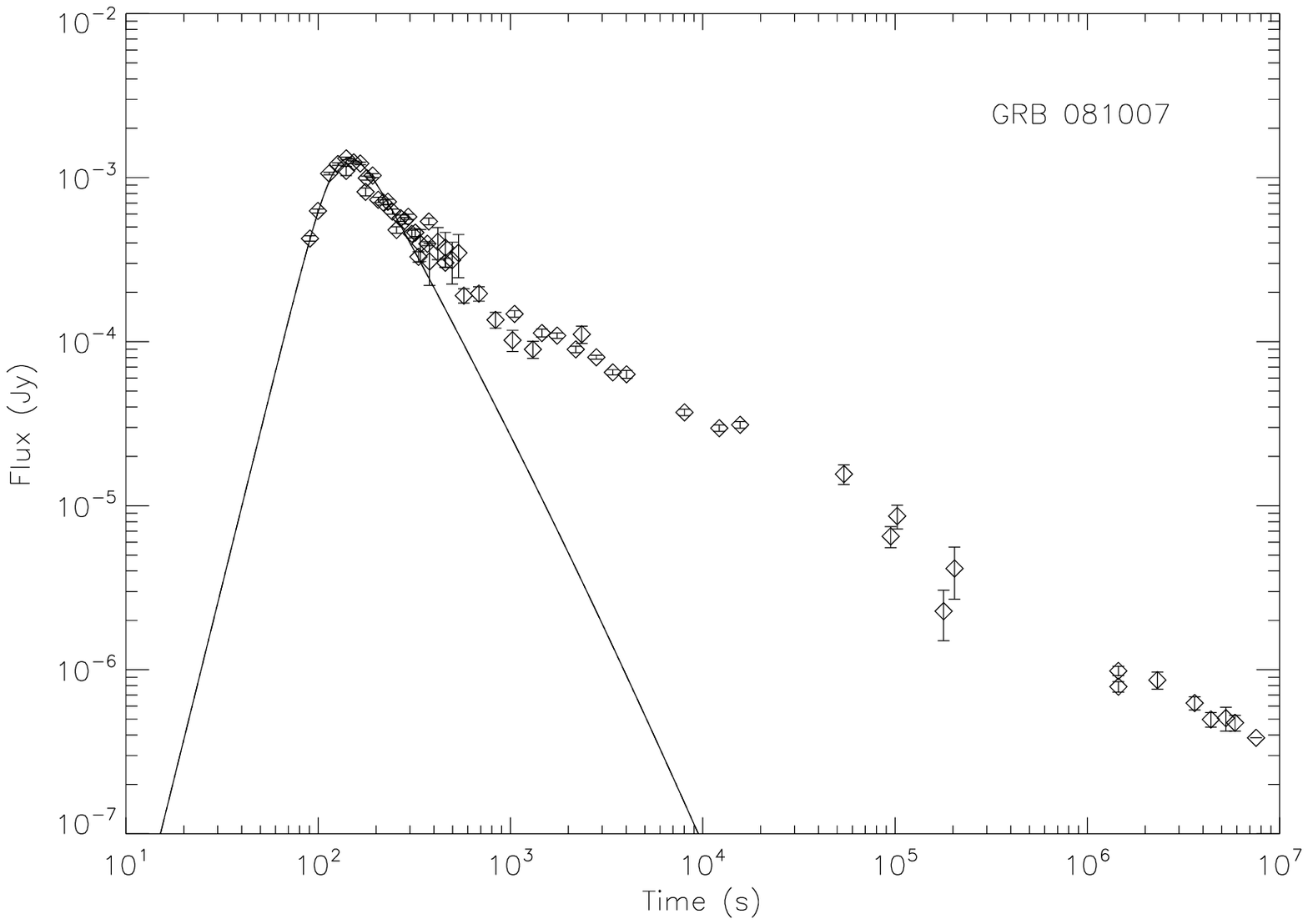}
\includegraphics[angle=0,scale=0.3]{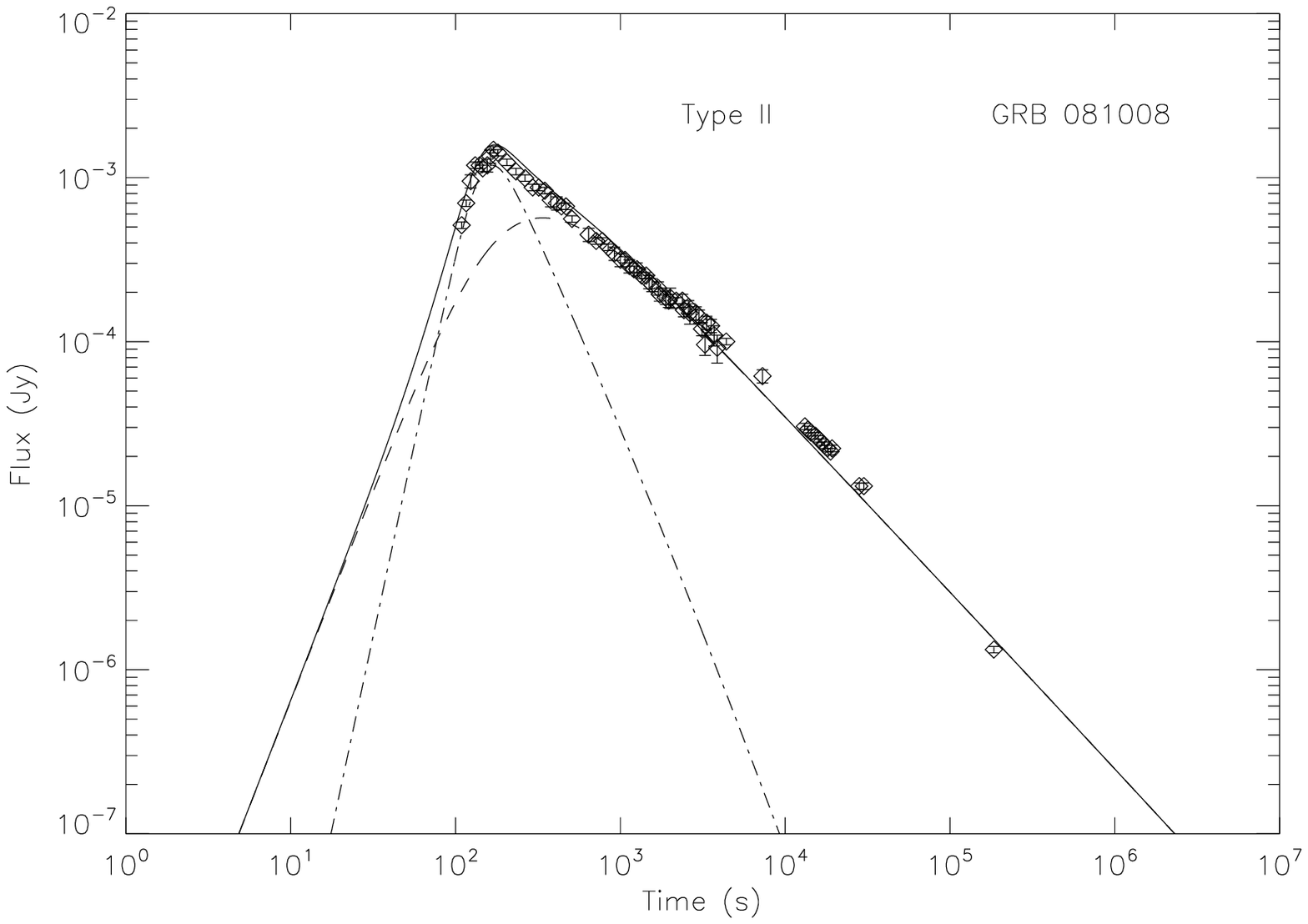}
\includegraphics[angle=0,scale=0.3]{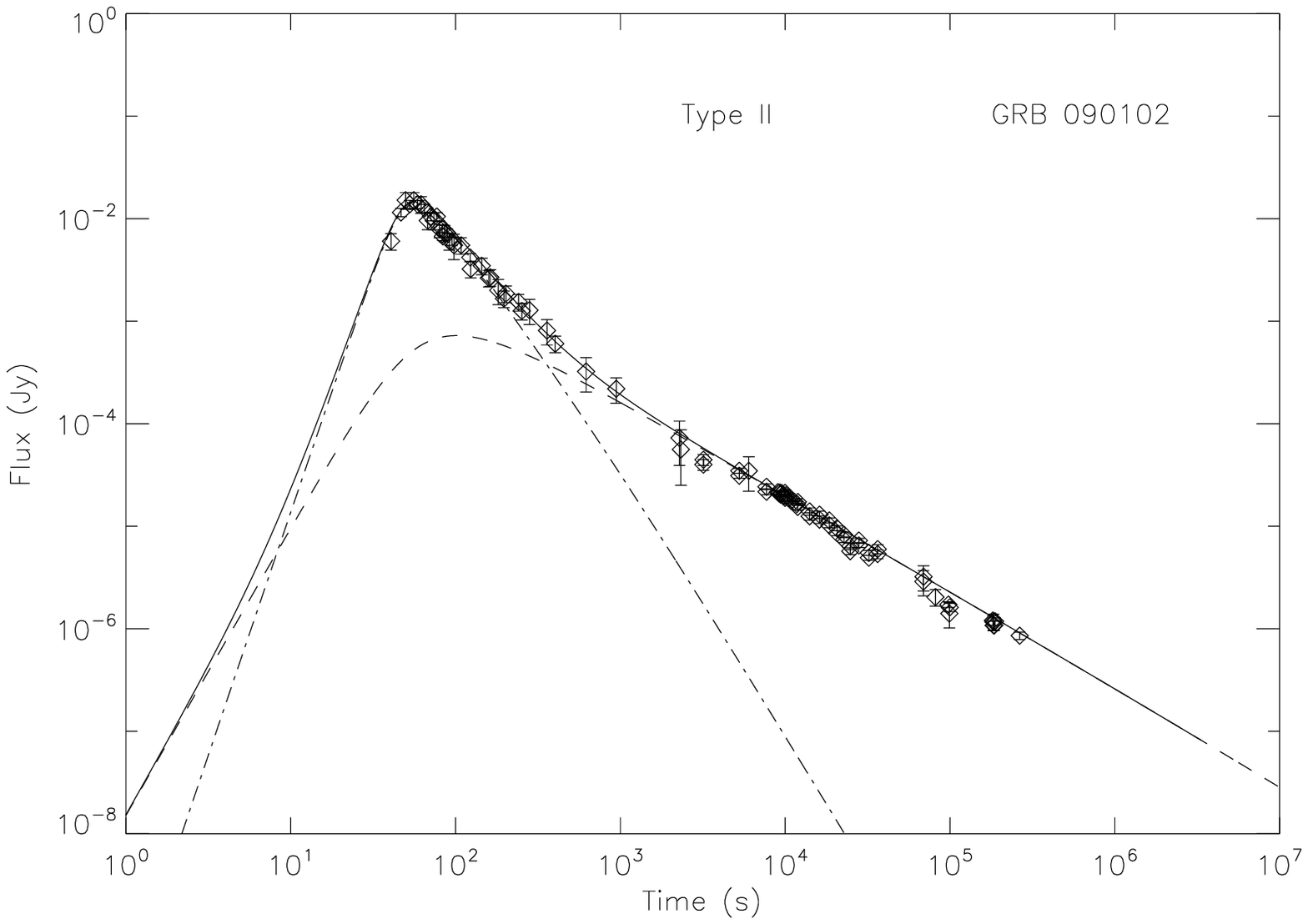}
\includegraphics[angle=0,scale=0.3]{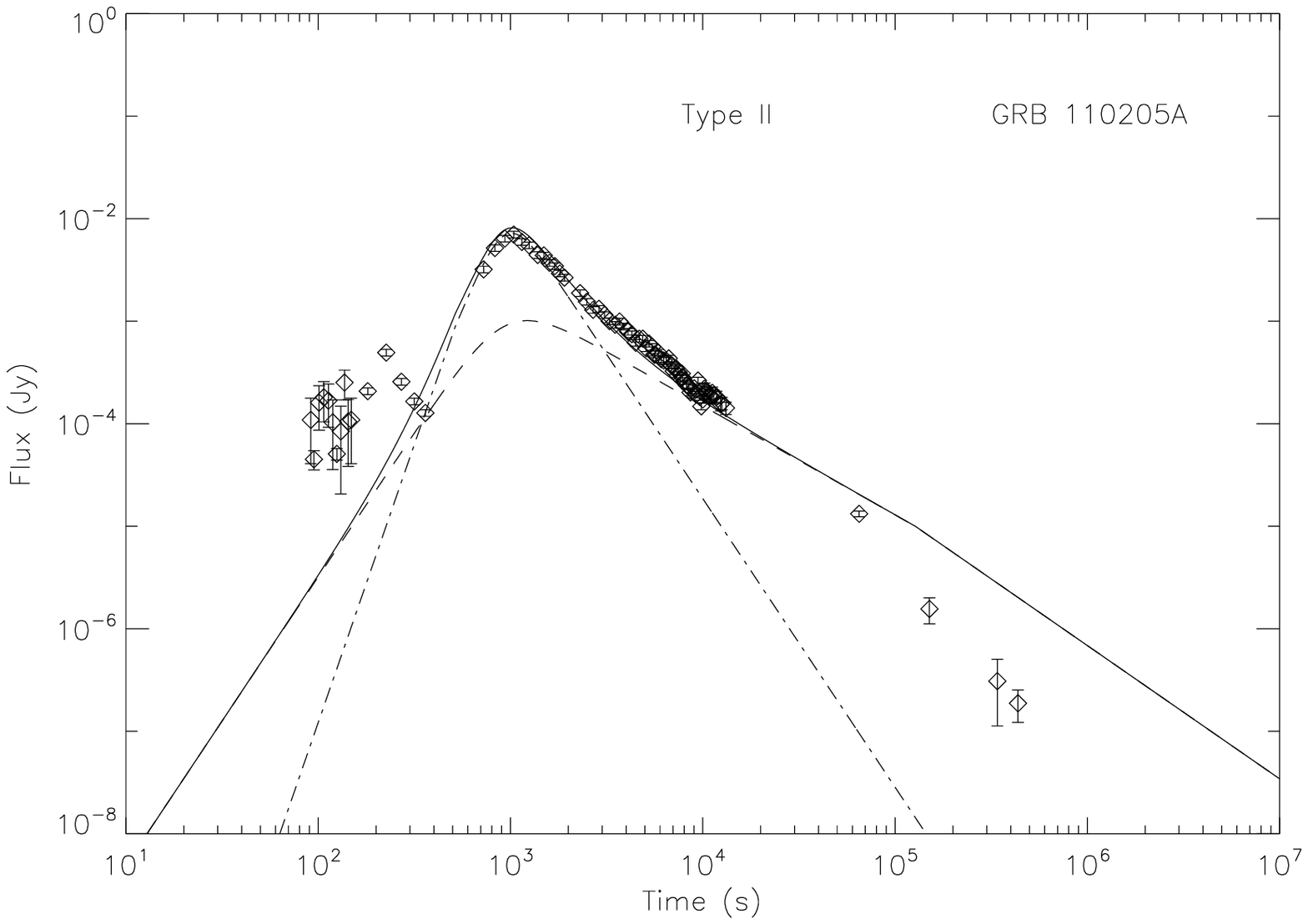}
\includegraphics[angle=0,scale=0.3]{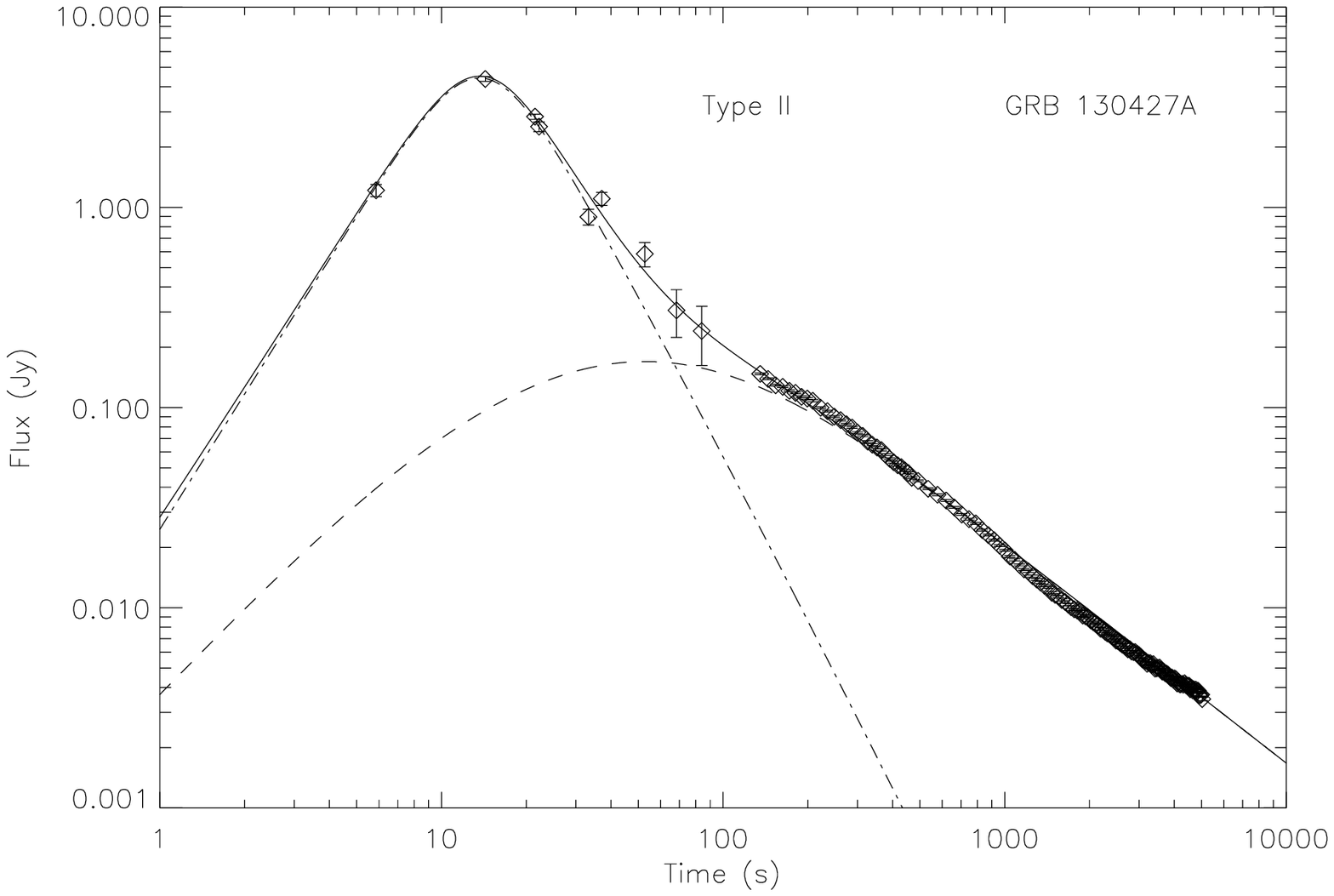}
\includegraphics[angle=0,scale=0.3]{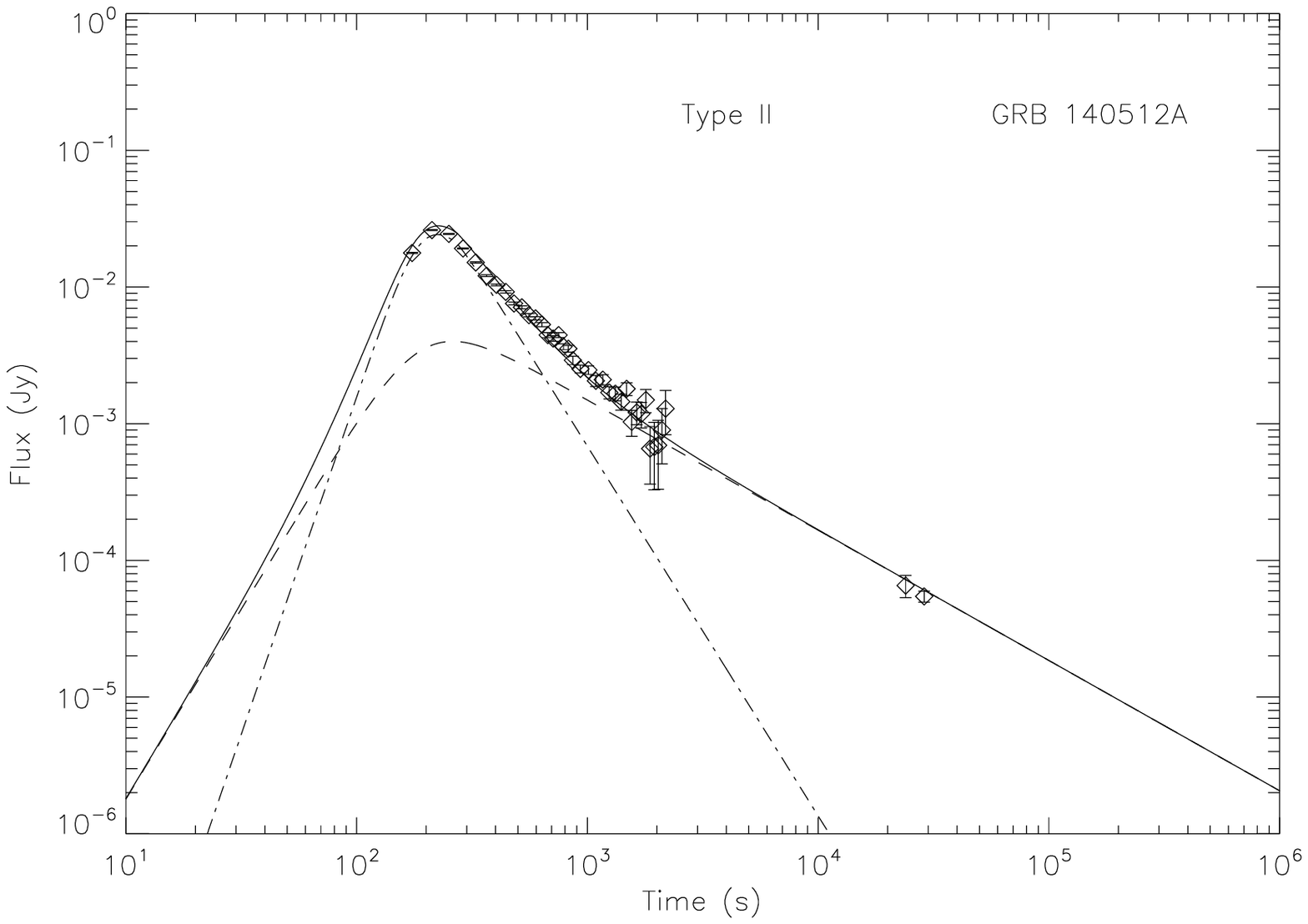}
\includegraphics[angle=0,scale=0.3]{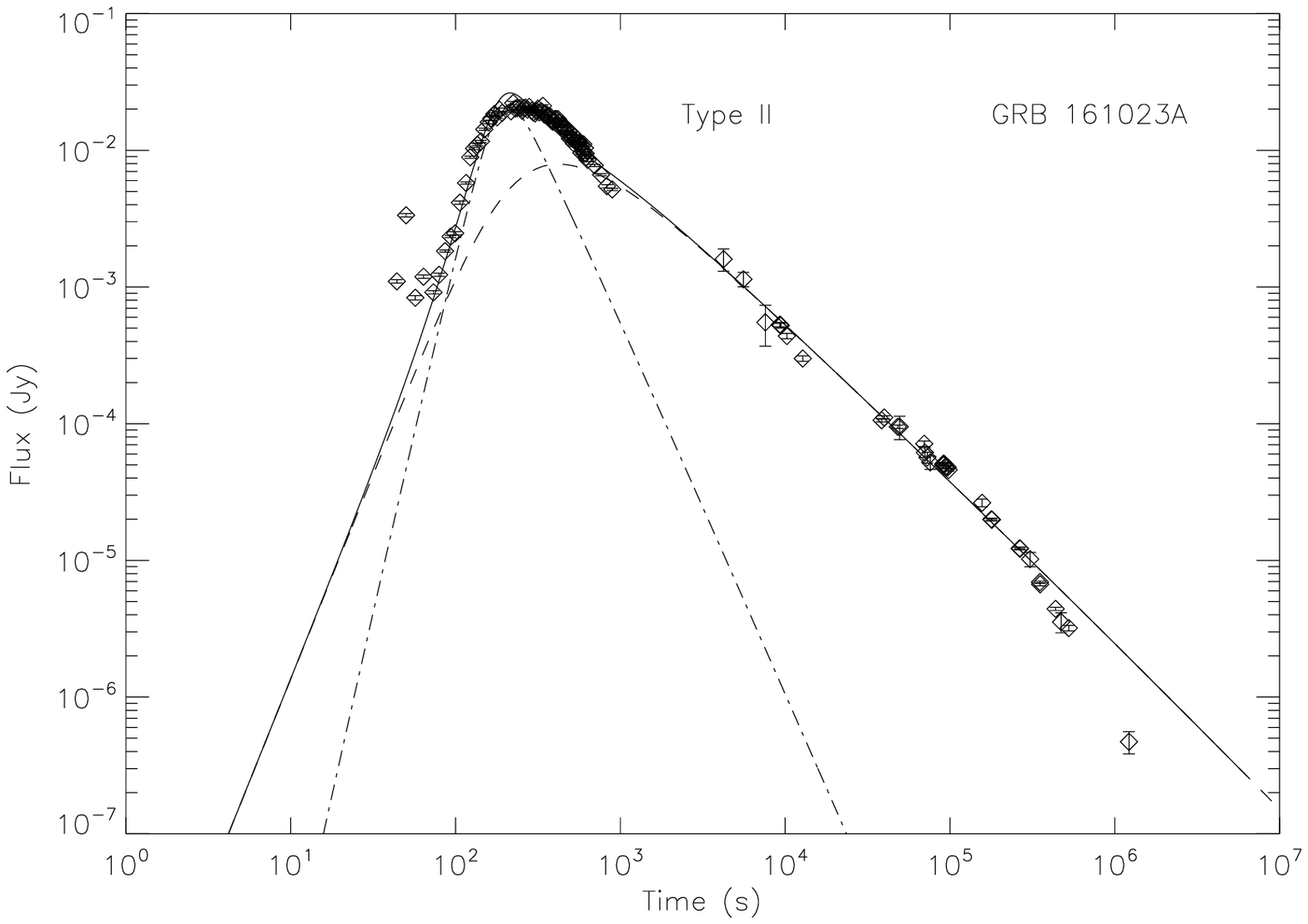}
\caption{The optical light curves dominated by the RS emission at the early time.
The smooth solid line is the fit to the optical data (diamonds), the dash-dotted
line marks the RS component, while the dashed line represents the FS component.}
\end{figure*}


\begin{figure*}
\includegraphics[angle=0,scale=0.35]{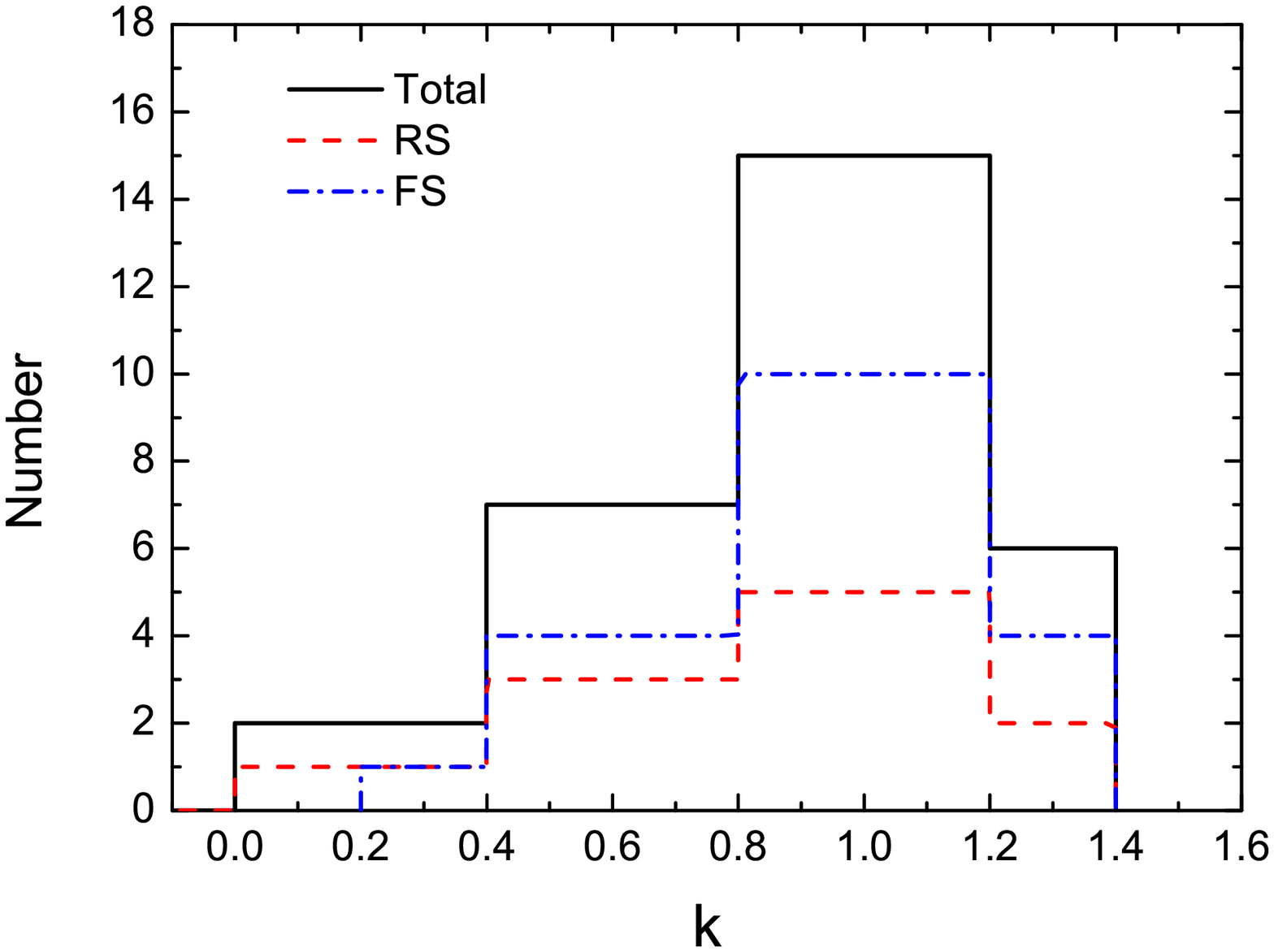}
\includegraphics[angle=0,scale=0.35]{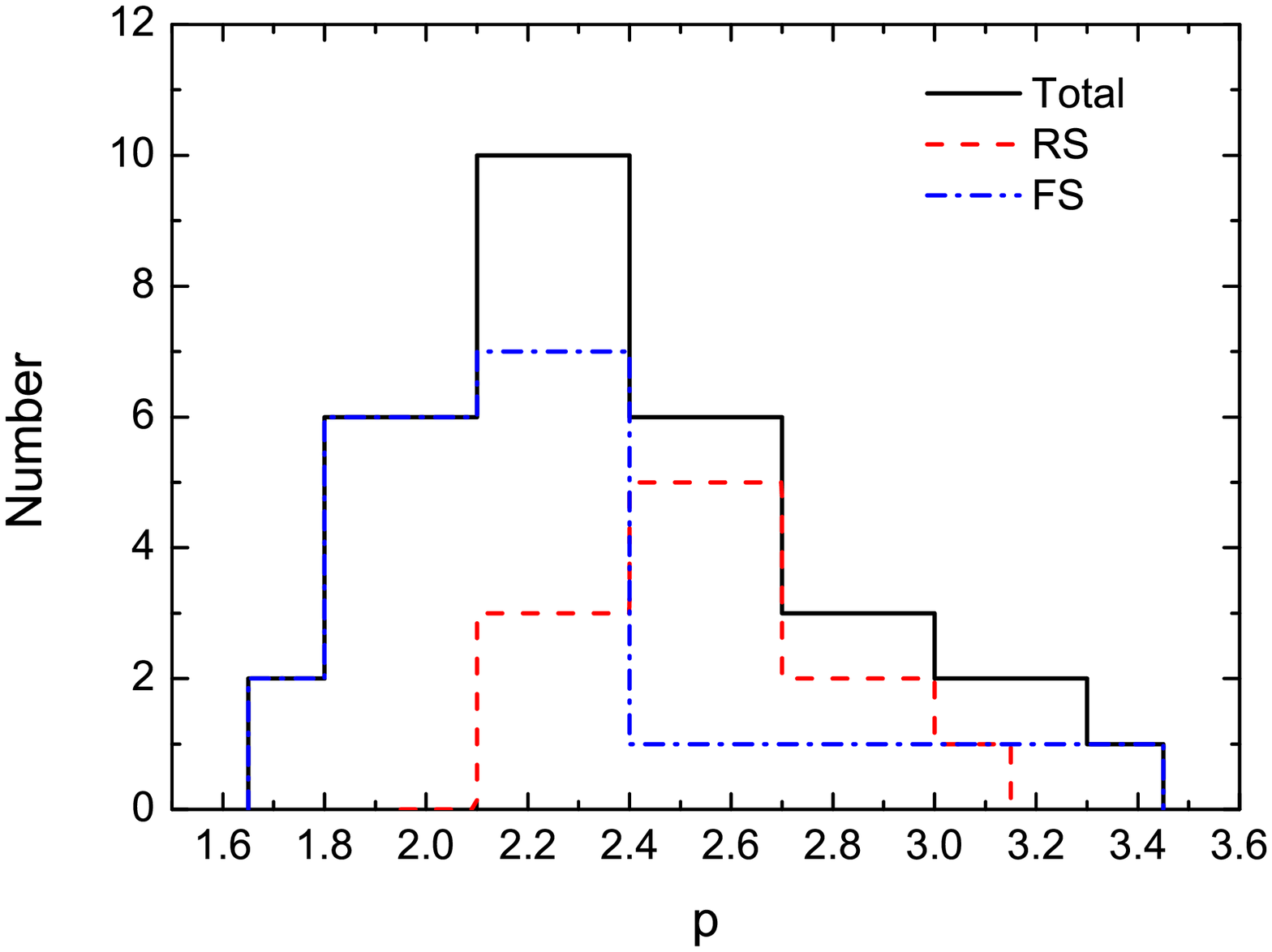}
\caption{Distributions of the values $k$ and $p$. The values $k$ and $p$ determined by the FS are taken from Yi et al. (2013). }
\end{figure*}

\clearpage
\begin{deluxetable}{ccccccccccccccccccccccccc}
\tabletypesize{\scriptsize}
\tablecaption{The fitting results for our optical sample. }
\tablewidth{0pt}

\tablehead{ \colhead{GRB}
&\colhead{$\beta_{o}$}
&\colhead{$E (10^{52}$ erg)}
&\colhead{$R_{0} (10^{17}$ cm)}
&\colhead{$\eta$}
&\colhead{$k$}
&\colhead{$p$}
&\colhead{$\varepsilon _{B, r}$}
&\colhead{$\varepsilon _{e, r}$}
&\colhead{$\varepsilon _{B, f}$}
&\colhead{$\varepsilon _{e, f}$}
&\colhead{Refs.}}

\startdata
990123	&	0.60	$\pm$	0.04	&	70	&	9	&	200	&	0.71	$\pm$	0.12	&	2.20	$\pm$	0.08	&	0.05	&	0.25	&	 0.002	 &	0.015	&	1	&\\
041219A	&	...			&	0.28	&	0.9	&	135	&	0.80	$\pm$	0.10	&	2.40			&	0.09	&	0.03	&	0.004	&	 0.045	 &	 2, 3	&\\
060607A	&	0.56	$\pm$	0.04	&	16	&	0.7	&	120	&	0.69	$\pm$	0.08	&	2.12	$\pm$	0.08	&	0.05	&	0.10	&	 0.003	 &	0.05	&	4	&\\
061007	&	0.78	$\pm$	0.02	&	350	&	1	&	160	&	0.76	$\pm$	0.03	&	2.56	$\pm$	0.04	&	0.04	&	0.06	&	 0.002	 &	0.12	&	5	&\\
081007	&	0.86	$\pm$	0.07	&	79	&	0.1	&	100	&	1.16	$\pm$	0.07	&	2.72	$\pm$	0.14	&	0.02	&	0.10	&	 ...	 &	...	&	6	&\\
081008	&	1.10	$\pm$	0.05	&	100	&	0.25	&	130	&	1.26	$\pm$	0.05	&	3.20	$\pm$	0.10	&	0.02	&	0.06	 &	 0.0015	&	0.075	&	7	&\\
090102	&	0.74	$\pm$	0.22	&	80	&	17	&	155	&	0.94	$\pm$	0.32	&	2.48	$\pm$	0.44	&	0.02	&	0.06	&	 5E-4	 &	0.052	&	8	&\\
110205A	&	0.84	$\pm$	0.01	&	90	&	0.5	&	80	&	0.86	$\pm$	0.03	&	2.68	$\pm$	0.02	&	0.06	&	0.03	&	 0.004	 &	0.006	&	9	&\\
130427A	&	0.70	$\pm$	0.05	&	400	&	40	&	155	&	1.48	$\pm$	0.06	&	2.40	$\pm$	0.10	&	0.02	&	0.06	&	 6E-4	 &	0.052	&	10	&\\
140521A	&	0.86	$\pm$	0.01	&	18	&	8	&	90	&	1.15	$\pm$	0.03	&	2.72	$\pm$	0.02	&	0.05	&	0.015	&	 0.0005	&	0.04	&	11	&\\
161023A	&	0.66	$\pm$	0.08	&	48	&	0.9	&	140	&	0.01	$\pm$	0.19	&	2.32	$\pm$	0.16	&	0.10	&	0.04	&	 0.01	 &	0.04	&	12	&\\
\enddata
\tablerefs{(1) Maiorano et al. 2005; (2) Blake et al. 2005; (3)
Fan et al. 2005; (4) Nysewander et al. 2009; (5) Zafar et al.
2011; (6) Japelj et al. 2014; (7) Yuan et al. 2010; (8) Gendre et al. 2010; (9) Gendre et al.
2012; (10) Vestrand et al. 2014; (11) Huang et al. 2016; (12) de Ugarte Postigo et al. 2018 }

\end{deluxetable}

\begin{deluxetable}{ccccccccccccccccccccccccc}
\tabletypesize{\scriptsize}
\tablecaption{The ratio of RS and FS parameters.}
\tablewidth{0pt}

\tablehead{ \colhead{GRB}
&\colhead{$R_{B}$}
&\colhead{$R_{e}$}
&\colhead{$\nu^{r}_{m,\Delta}/\nu^{f}_{m,\Delta}$}
&\colhead{$\nu^{r}_{c,\Delta}/\nu^{f}_{c,\Delta}$}
&\colhead{$F_{\nu ,\max, \Delta }^{RS}/F_{\nu ,\max, \Delta }^{FS}$}}

\startdata
990123	&	5.1	&	16.7	&	0.012	&	0.007	&	2562.5	&\\
041219A	&	4.8	&	0.6	&	1.622E-5	&	0.009	&	1850.9	&\\
060607A	&	4.2	&	2.0	&	3.485E-4	&	0.013	&	1301.9	&\\
061007	&	4.7	&	0.5	&	8.939E-6	&	0.010	&	2129.1	&\\
081008	&	4.0	&	0.8	&	1.263E-5	&	0.016	&	1824.5	&\\
090102	&	6.9	&	1.2	&	8.292E-5	&	0.003	&	2970.4	&\\
110205A	&	4.0	&	4.2	&	0.002	&	0.016	&	913.1	&\\
130427A	&	6.1	&	1.1	&	1.933E-5	&	0.005	&	3521.2	&\\
140521A	&	10.2	&	0.4	&	1.68E-5	&	0.001	&	3076.4	&\\
161023A	&	3.2	&	1.1	&	1.2E-4	&	0.032	&	920.5	&\\
\enddata

\end{deluxetable}


\begin{thebibliography}{}
\bibitem[Ackermann et al.(2014)]{2014Sci...343...42A} Ackermann, M.,
Ajello, M., Asano, K., et al.\ 2014, Science, 343, 42


\bibitem[Amaro-Seoane et al.(2017)]{2017arXiv1702.00786J} Amaro-Seoane, P., Audley, H., Babak, S., et al.\ 2017, arXiv:1702.00786


\bibitem[Blake et al.(2005)]{2005Natur.435..181B} Blake, C.~H., Bloom,
J.~S., Starr, D.~L., et al.\ 2005, \nat, 435, 181

\bibitem[\protect\citeauthoryear{Burrows et al.}{2005}]{2005SSRv..120..165B} Burrows D.~N., et al., 2005a, SSRv, 120, 165



\bibitem[Burrows et al.(2005)]{2005Sci...309.1833B} Burrows, D.~N., Romano,
P., Falcone, A., et al.\ 2005b, Science, 309, 1833


\bibitem[Castro-Tirado et al.(1999)]{1999Sci...283.2069C} Castro-Tirado,
A.~J., Zapatero-Osorio, M.~R., Caon, N., et al.\ 1999, Science, 283, 2069


\bibitem[Chevalier
\& Li(2000)]{2000ApJ...536..195C} Chevalier, R.~A., \& Li, Z.-Y.\ 2000, \apj, 536, 195


\bibitem[Covino et al.(2008)]{2008ChJAS...8..356C} Covino, S., Vergani,
S.~D., Malesani, D., et al.\ 2008, Chinese Journal of Astronomy and
Astrophysics Supplement, 8, 356

\bibitem[Dai
\& Lu(1998)]{1998A&A...333L..87D} Dai, Z.~G., \& Lu, T.\ 1998a, \aap, 333, L87

\bibitem[Dai
\& Lu(1998)]{1998PhRvL..81.4301D} Dai, Z.~G., \& Lu, T.\ 1998b, Physical Review Letters, 81, 4301


\bibitem[Dai
\& Lu(1998)]{1998MNRAS.298...87D} Dai, Z.~G., \& Lu, T.\ 1998c, \mnras, 298, 87

\bibitem[Dai
\& Lu(1999)]{1999ApJ...519L.155D} Dai, Z.~G., \& Lu, T.\ 1999, \apjl, 519, L155


\bibitem[Dai(2004)]{2004ApJ...606.1000D} Dai, Z.~G.\ 2004, \apj, 606, 1000


\bibitem[Dai et al.(2006)]{2006Sci...311.1127D} Dai, Z.~G., Wang, X.~Y.,
Wu, X.~F., \& Zhang, B.\ 2006, Science, 311, 1127

\bibitem[De Colle et al.(2012)]{2012ApJ...751...57D} De Colle, F., Ramirez-Ruiz, E., Granot, J., \& Lopez-Camara, D.\ 2012, \apj, 751, 57


\bibitem[de Ugarte Postigo et al.(2018)]{2018A&A...620A.119D} de Ugarte Postigo, A., Th{\"o}ne, C.~C., Bolmer, J., et al.\ 2018, \aap, 620, A119

\bibitem[\protect\citeauthoryear{Fan, et al.}{2002}]{2002ChJAA...2..449F} Fan Y.-Z., Dai Z.-G., Huang Y.-F., Lu T., 2002, ChJAA, 2, 449

\bibitem[Fan et al.(2005)]{2005ApJ...628L..25F} Fan, Y.~Z., Zhang, B.,\& Wei, D.~M.\ 2005, \apjl, 628, L25


\bibitem[Fan et al.(2013)]{2013ApJ...776...95F} Fan, Y.-Z., Tam, P.~H.~T.,
Zhang, F.-W., et al.\ 2013, \apj, 776, 95


\bibitem[G{\"o}tz et al.(2011)]{2011MNRAS.413.2173G} G{\"o}tz, D., Covino,
S., Hasco{\"e}t, R., et al.\ 2011, \mnras, 413, 2173


\bibitem[\protect\citeauthoryear{Gao, et al.}{2013}]{2013NewAR..57..141G} Gao H., Lei W.-H., Zou Y.-C., Wu X.-F., Zhang B., 2013, NewAR, 57, 141

\bibitem[Gao et al.(2015)]{2015ApJ...810..160G} Gao, H., Wang, X.-G., M{\'e}sz{\'a}ros, P., \& Zhang, B.\ 2015, \apj, 810, 160

\bibitem[\protect\citeauthoryear{Gehrels et al.}{2004}]{2004ApJ...611.1005G} Gehrels N., et al., 2004, ApJ, 611, 1005


\bibitem[Gendre et al.(2012)]{2012ApJ...748...59G} Gendre, B., Atteia,
J.~L., Bo{\"e}r, M., et al.\ 2012, \apj, 748, 59


\bibitem[Gendre et al.(2010)]{2010MNRAS.405.2372G} Gendre, B., Klotz, A.,
Palazzi, E., et al.\ 2010, \mnras, 405, 2372

\bibitem[Giannios et al.(2008)]{2008A&A...478..747G} Giannios, D., Mimica, P., \& Aloy, M.~A.\ 2008, \aap, 478, 747


\bibitem[Gomboc et al.(2009)]{2009AIPC.1133..145G} Gomboc, A., Kobayashi,
S., Mundell, C.~G., et al.\ 2009, American Institute of Physics Conference
Series, 1133, 145

\bibitem[Granot, \& Sari(2002)]{2002ApJ...568..820G} Granot, J., \& Sari, R.\ 2002, \apj, 568, 820



\bibitem[Harrison
\& Kobayashi(2013)]{2013ApJ...772..101H} Harrison, R., \& Kobayashi, S.\ 2013, \apj, 772, 101

\bibitem[Higgins et al.(2019)]{2019MNRAS.484.5245H} Higgins, A.~B., van der Horst, A.~J., Starling, R.~L.~C., et al.\ 2019, \mnras, 484, 5245



\bibitem[Hou et al.(2014)]{2014ApJ...785..113H} Hou, S.~J., Geng, J.~J., Wang, K., et al.\ 2014, \apj, 785, 113



\bibitem[Huang et al.(2016)]{2016ApJ...833..100H} Huang, X.-L., Xin, L.-P., Yi, S.-X., et al.\ 2016, \apj, 833, 100



\bibitem[Japelj et al.(2014)]{2014ApJ...785...84J} Japelj, J., Kopa{\v c}, D., Kobayashi, S., et al.\ 2014, \apj, 785, 84


\bibitem[Jin
\& Fan(2007)]{2007MNRAS.378.1043J} Jin, Z.~P., \& Fan, Y.~Z.\ 2007, \mnras, 378, 1043


\bibitem[Jin et al.(2013)]{2013ApJ...774..114J} Jin, Z.-P., Covino, S.,
Della Valle, M., et al.\ 2013, \apj, 774, 114

\bibitem[Kann et al.(2010)]{2010ApJ...720.1513K} Kann, D.~A., Klose, S., Zhang, B., et al.\ 2010, \apj, 720, 1513

\bibitem[Kawamura et al.(2006)]{2006CQGra..23S.125K} Kawamura, S., Nakamura, T., Ando, M., et al.\ 2006, Classical and Quantum Gravity, 23, S125


\bibitem[Klotz et al.(2009)]{2009GCN..8764....1K} Klotz, A., Gendre, B.,
Boer, M., \& Atteia, J.~L.\ 2009, GRB Coordinates Network, 8764, 1


\bibitem[Kobayashi(2000)]{2000ApJ...545..807K} Kobayashi, S.\ 2000, \apj,
545, 807


\bibitem[Kobayashi
\& Zhang(2003)]{2003ApJ...582L..75K} Kobayashi, S., \& Zhang, B.\ 2003, \apjl, 582, L75


\bibitem[Kouveliotou et al.(2013)]{2013ApJ...779L...1K} Kouveliotou, C.,
Granot, J., Racusin, J.~L., et al.\ 2013, \apjl, 779, L1

\bibitem[Lamb et al.(2019)]{2019ApJ...883...48L} Lamb, G.~P., Tanvir, N.~R., Levan, A.~J., et al.\ 2019, \apj, 883, 48

\bibitem[L{\"u} \& Zhang(2014)]{2014ApJ...785...74L} L{\"u}, H.-J., \& Zhang, B.\ 2014, \apj, 785, 74


\bibitem[Laskar et al.(2013)]{2013ApJ...776..119L} Laskar, T., Berger, E.,
Zauderer, B.~A., et al.\ 2013, \apj, 776, 119


\bibitem[Liang et al.(2013)]{2013ApJ...774...13L} Liang, E.-W., Li, L.,
Gao, H., et al.\ 2013, \apj, 774, 13


\bibitem[Liang et al.(2010)]{2010ApJ...725.2209L} Liang, E.-W., Yi, S.-X.,
Zhang, J., et al.\ 2010, \apj, 725, 2209


\bibitem[Liu et al.(2013)]{2013ApJ...773L..20L} Liu, R.-Y., Wang, X.-Y.,
\& Wu, X.-F.\ 2013, \apjl, 773, L20

\bibitem[Lloyd-Ronning(2018)]{2018Galax...6..103L} Lloyd-Ronning, N.\ 2018, Galaxies, 6, 103
\bibitem[Lloyd-Ronning, \& Fryer(2017)]{2017MNRAS.467.3413L} Lloyd-Ronning, N.~M., \& Fryer, C.~L.\ 2017, \mnras, 467, 3413


\bibitem[Luo et al.(2016)]{2016CQGra..33c5010L} Luo, J., Chen, L.-S., Duan, H.-Z., et al.\ 2016, Classical and Quantum Gravity, 33, 035010



\bibitem[Maiorano et
al.(2005)]{2005A&A...438..821M} Maiorano, E., Masetti, N., Palazzi, E., et al.\ 2005, \aap, 438, 821


\bibitem[Maselli et al.(2014)]{2014Sci...343...48M} Maselli, A., Melandri,
A., Nava, L., et al.\ 2014, Science, 343, 48


\bibitem[M{\'e}sz{\'a}ros
\& Rees(1997)]{1997ApJ...476..232M} M{\'e}sz{\'a}ros, P., \& Rees, M.~J.\ 1997, \apj, 476, 232

\bibitem[M{\'e}sz{\'a}ros
\& Rees(1999)]{1999MNRAS.306L..39M} M{\'e}sz{\'a}ros, P., \& Rees, M.~J.\ 1999, \mnras, 306, L39

\bibitem[Molinari et
al.(2007)]{2007A&A...469L..13M} Molinari, E., Vergani, S.~D., Malesani, D., et al.\ 2007, \aap, 469, L13

\bibitem[Nakar
\& Piran(2005)]{2005ApJ...619L.147N} Nakar, E., \& Piran, T.\ 2005, \apjl, 619, L147


\bibitem[Nousek et al.(2006)]{2006ApJ...642..389N} Nousek, J.~A.,
Kouveliotou, C., Grupe, D., et al.\ 2006, \apj, 642, 389


\bibitem[Nysewander et al.(2009)]{2009ApJ...693.1417N} Nysewander, M.,
Reichart, D.~E., Crain, J.~A., et al.\ 2009, \apj, 693, 1417


\bibitem[Panaitescu
\& Kumar(2004)]{2004MNRAS.353..511P} Panaitescu, A., \& Kumar, P.\ 2004, \mnras, 353, 511

\bibitem[Panaitescu, \& Vestrand(2011)]{2011MNRAS.414.3537P} Panaitescu, A., \& Vestrand, W.~T.\ 2011, \mnras, 414, 3537


\bibitem[Panaitescu et al.(2013)]{2013MNRAS.436.3106P} Panaitescu, A.,
Vestrand, W.~T., \& Wo{\'z}niak, P.\ 2013, \mnras, 436, 3106


\bibitem[Perley et al.(2014)]{2014ApJ...781...37P} Perley, D.~A., Cenko,
S.~B., Corsi, A., et al.\ 2014, \apj, 781, 37


\bibitem[Piran(1999)]{1999PhR...314..575P} Piran, T.\ 1999, \physrep, 314,
575

\bibitem[Roming et al.(2005)]{2005SSRv..120...95R} Roming, P.~W.~A., Kennedy, T.~E., Mason, K.~O., et al.\ 2005, \ssr, 120, 95


\bibitem[\protect\citeauthoryear{Rueda, et al.}{2018}]{2018JCAP...10..006R} Rueda J.~A., et al., 2018, JCAP, 2018, 006

\bibitem[Rykoff et al.(2009)]{2009ApJ...702..489R} Rykoff, E.~S., Aharonian, F., Akerlof, C.~W., et al.\ 2009, \apj, 702, 489




\bibitem[Sari
\& Piran(1999)]{1999ApJ...520..641S} Sari, R., \& Piran, T.\ 1999a, \apj, 520, 641


\bibitem[Sari
\& Piran(1999)]{1999ApJ...517L.109S} Sari, R., \& Piran, T.\ 1999b, \apjl, 517, L109


\bibitem[Sari
\& Piran(1995)]{1995ApJ...455L.143S} Sari, R., \& Piran, T.\ 1995, \apjl, 455, L143


\bibitem[Sari et al.(1998)]{1998ApJ...497L..17S} Sari, R., Piran, T.,
\& Narayan, R.\ 1998, \apjl, 497, L17


\bibitem[Schulze et
al.(2011)]{2011A&A...526A..23S} Schulze, S., Klose, S., Bj{\"o}rnsson, G., et al.\ 2011, \aap, 526, A23

\bibitem[Si et al.(2018)]{2018ApJ...863...50S} Si, S.-K., Qi, Y.-Q., Xue, F.-X., et al.\ 2018, \apj, 863, 50


\bibitem[Tam et al.(2013)]{2013ApJ...771L..13T} Tam, P.-H.~T., Tang, Q.-W.,
Hou, S.-J., Liu, R.-Y., \& Wang, X.-Y.\ 2013, \apjl, 771, L13


\bibitem[Vestrand et al.(2014)]{2014Sci...343...38V} Vestrand, W.~T., Wren,
J.~A., Panaitescu, A., et al.\ 2014, Science, 343, 38


\bibitem[Wang
\& Dai(2013)]{2013NatPh...9..465W} Wang, F.~Y., \& Dai, Z.~G.\ 2013, Nature Physics, 9, 465


\bibitem[Wang et al.(2000)]{2000MNRAS.319.1159W} Wang, X.~Y., Dai, Z.~G.,
\& Lu, T.\ 2000, \mnras, 319, 1159


\bibitem[Wijers et al.(1997)]{1997MNRAS.288L..51W} Wijers, R.~A.~M.~J.,
Rees, M.~J., \& Meszaros, P.\ 1997, \mnras, 288, L51


\bibitem[Wu et al.(2005)]{2005ApJ...619..968W} Wu, X.~F., Dai, Z.~G.,
Huang, Y.~F., \& Lu, T.\ 2005, \apj, 619, 968


\bibitem[Wu et al.(2003)]{2003MNRAS.342.1131W} Wu, X.~F., Dai, Z.~G.,
Huang, Y.~F., \& Lu, T.\ 2003, \mnras, 342, 1131


\bibitem[Wu et al.(2013)]{2013ApJ...767L..36W} Wu, X.-F., Hou, S.-J.,
\& Lei, W.-H.\ 2013, \apjl, 767, L36


\bibitem[Xu et al.(2013)]{2013ApJ...776...98X} Xu, D., de Ugarte Postigo,
A., Leloudas, G., et al.\ 2013, \apj, 776, 98

\bibitem[Yi et al.(2013)]{2013ApJ...776..120Y} Yi, S.-X., Wu, X.-F.,
\& Dai, Z.-G.\ 2013, \apj, 776, 120

\bibitem[Yi et al.(2014)]{2014ApJ...792L..21Y} Yi, S.-X., Gao, H.,
\& Zhang, B.\ 2014, \apjl, 792, LL21

\bibitem[Yi et al.(2017)]{2017ApJ...844...79Y} Yi, S.-X., Yu, H., Wang, F.~Y., \& Dai, Z.-G.\ 2017, \apj, 844, 79


\bibitem[Yi et al.(2016)]{2016ApJS..224...20Y} Yi, S.-X., Xi, S.-Q., Yu, H., et al.\ 2016, \apjs, 224, 20


\bibitem[Yi et al.(2015)]{2015ApJ...807...92Y} Yi, S.-X., Wu, X.-F., Wang,
F.-Y., \& Dai, Z.-G.\ 2015, \apj, 807, 92

\bibitem[Yuan et al.(2010)]{2010ApJ...711..870Y} Yuan, F., Schady, P.,
Racusin, J.~L., et al.\ 2010, \apj, 711, 870


\bibitem[Zafar et
al.(2011)]{2011A&A...532A.143Z} Zafar, T., Watson, D., Fynbo, J.~P.~U., et al.\ 2011, \aap, 532, A143


\bibitem[Zhang(2013)]{2013ApJ...763L..22Z} Zhang, B.\ 2013, \apjl, 763, L22


\bibitem[\protect\citeauthoryear{Zhang}{2007}]{2007ChJAA...7....1Z} Zhang B., 2007, ChJAA, 7, 1


\bibitem[Zhang et al.(2006)]{2006ApJ...642..354Z} Zhang, B., Fan, Y.~Z.,
Dyks, J., et al.\ 2006, \apj, 642, 354


\bibitem[Zhang
\& Kobayashi(2005)]{2005ApJ...628..315Z} Zhang, B., \& Kobayashi, S.\ 2005, \apj, 628, 315


\bibitem[Zhang et al.(2003)]{2003ApJ...595..950Z} Zhang, B., Kobayashi, S.,
\& M{\'e}sz{\'a}ros, P.\ 2003, \apj, 595, 950


\bibitem[Zhang \& M{\'e}sz{\'a}ros(2001)]{2001ApJ...552L..35Z} Zhang, B., \& M{\'e}sz{\'a}ros, P.\ 2001, \apjl, 552, L35

\bibitem[Zhou et al.(2020)]{2020IJMPD} Zhou, Q.-Q., Yi, S.-X., Huang, X.-L., et al.\ 2020, International Journal of Modern Physics D, accepted


\bibitem[Ziaeepour et al.(2008)]{2008MNRAS.385..453Z} Ziaeepour, H.,
Holland, S.~T., Boyd, P.~T., et al.\ 2008, \mnras, 385, 453


\bibitem[Zou et al.(2005)]{2005MNRAS.363...93Z} Zou, Y.~C., Wu, X.~F.,
\& Dai, Z.~G.\ 2005, \mnras, 363, 93



\end{thebibliography}
\end{document}